\documentclass[aps,prd,preprint,nofootinbib,groupedaddress]{revtex4-1}

\usepackage{amsfonts,amssymb,amsmath}
\usepackage{graphicx}
\usepackage{color}
\usepackage{hyperref}

\usepackage{todonotes}

\usepackage[framemethod=TikZ]{mdframed}
\usepackage{lipsum}

\mdfdefinestyle{MyFrame}{
linecolor=blue,
outerlinewidth=2pt,
roundcorner=20pt,
innertopmargin=\baselineskip,
innerbottommargin=\baselineskip,
innerrightmargin=20pt,
innerleftmargin=20pt,
backgroundcolor=gray!20!white}

\newcommand{\be}{\begin{equation}}
\newcommand{\ee}{\end{equation}}
\newcommand{\bea}{\begin{eqnarray}}
\newcommand{\eea}{\end{eqnarray}}
\newcommand{\beal}{\begin{aligned}}
\newcommand{\eeal}{\end{aligned}}

\usepackage{color}

\def\bk{\hfill\break}
\def\K{ {\cal K} }

\def\pbh{ {\cal H}_b^- }
\def\fbh{ {\cal H}_b^+ }
\def\pch{ {\cal H}_c^- }
\def\fch{ {\cal H}_c^+ }

\def\omegap{\omega^\prime}
\def\betab{\beta^b_{\omega \omega'} }
\def\betac{\beta^c_{\omega \omega'} }
\def\alphab{\alpha^b_{\omega \omega'} }
\def\alphac{\alpha^c_{\omega \omega'} }
\def\kb{\kappa_b} 
\def\kc{\kappa_c}
\def\kt{\kappa_T}

\def\e{\epsilon}

\def\be{\begin{equation}}
\def\ee{\end{equation}}
\def\bea{\begin{eqnarray}}
\def\eea{\end{eqnarray}}
\def\ba{\begin{align}}
\def\ea{\end{align}}

\begin{document}

\begin{titlepage}
\vfill

\begin{center}
\baselineskip=16pt
{\Large\bf Black Hole and Cosmological Particle Production 
in Schwarzschild de Sitter}

\vskip 2.mm

{\bf   Yue Qiu and Jennie Traschen} 

\vskip 1.mm

Department of Physics\\ University of Massachusetts, Amherst, MA 01003, USA\\

\vskip 0.1 in Email: \texttt{yqiu@physics.umass.edu, traschen@physics.umass.edu}
\vspace{6pt}
\end{center}
\vskip 0.3in
\par
\begin{center}
{\bf Abstract}
\end{center}
\begin{quote}

We compute the spectra and total fluxes of quantum mechanically produced particles crossing
the black hole and cosmological horizons in Schwarzchild de Sitter (SdS). 
Particle states are defined with respect to well-behaved, Kruskal coordinates 
near the horizons, and as a consequence we find that these spectra are generally
non-thermal. 
The non-thermal Bogoliubov coefficient for a vacuum fluctuation near the black hole horizon to produce a particle that crosses the cosmological horizon is shown to equal to the convolution
of two thermal coefficients, one at the cosmological temperature and one at the black hole temperature, 
weighted by the transmission coefficient for wave propagation in static SdS coordinates. In this sense virtual
thermal propagation underlies the production process. This representation
leads to the useful result that the geometric optics approximation is reliable when used together with 
a low frequency cut-off determined by the transmission coefficient. The large black hole limit is a quasi-equilibruim
situation as both temperatures approach the common value of zero, the particle spectra become equal, and 
both emissions are exponentially suppressed. Small black holes radiate as thermal bodies and absorb
a tiny flux of cosmological particles. The behavior of the quantum fluctuations on the horizons
is seen to be consistent with the Schottky anomaly behavior of classical
gravitational fluctuations.

\end{quote}
\end{titlepage}

\section{Introduction}

Hawking taught us  that the formal identification of the surface gravity $\kb$ of a black hole with a temperature, as suggested
by the first law of black hole mechanics \cite{Bardeen:1973gs}, is 
realized quantum mechanically \cite{Hawking:1974sw}. Considering an asymptotically flat black hole 
that is initially in the vacuum state of a quantum field, he calculated that far from the black hole and at late times
 there is a flux of quantum mechanical particles. Famously, the spectrum of the 
radiation is thermal with temperature $ 2\pi T_b = \kappa_b$. Subsequently reference \cite{Gibbons:1977mu}
showed that the cosmological horizon in de Sitter also leads to the thermal emission of quantum paricles
with temperature $ 2\pi T_c = \kappa_c$, where is the magnitude of the cosmological horizon surface gravity.   

In a black hole spacetime with positive $\Lambda $ there is particle production due to both the black hole and 
cosmological horizons. This is an intriguing situation, since 
the two surface gravities are not typically equal, which is at odds with our usual expectations of thermal equilibrium. Of course,
 unequal temperatures is also a feature of asymptotically flat black holes if one assigns $T_\infty =0$
to the asymptotically flat region, and is consistent with the fact that black holes evaporate.
Thermal properties of Schwarzschild-de Sitter (SdS) black holes were explored in \cite{Gibbons:1977mu} using general considerations.
Particle production for charged de Sitter black holes was computed in \cite{Kastor:1993mj} which focused on the 
the special case of $|Q| =M$ when both temperatures are equal and nonzero. It was found that the spectra for
$|Q| =M$ black holes is not thermal. In this paper we greatly expand on the work of \cite{Kastor:1993mj}. We
calculate the particle spectra and the total particle production rates for
 both the black hole and cosmological horizons  in  Schwarzschild-de Sitter. 
With the exception of radiation from small black holes, we find that these spectra are non-thermal. This
perhaps unexpected result arises from our 
choice of particle states. To compute particle fluxes crossing the horizons, either being absorbed by the black hole
or flowing with the expanding universe across the cosmiological horizon, it is necessary to use modes for
 the particle states that are well-behaved on the horizons. The natural choice for the time coordinate is to use
 an affine parameter along null geodesics, that is,
Kruskal coordinates. This choice constitutes a generalization of the Unruh vacuum in Schwarzschild
 \cite{Unruh:1976db} to SdS \cite{Balbinot:1985mk}. 
 
 The motivation for interest in particles that are well defined on the horizons was, in part, to explore quantum aspects of the
 Schottky anomaly of SdS black holes, discussed in a companion paper
 \cite{Dinsmore:2019elr}.  Schottky behavior refers to peaks in the specific heat $dE/dT$ and in $dS/dT$
 for certain statistical mechanics systems. In SdS one finds that there is a Schottky-type peak in $dS_{tot}/dT_b $, where
 $S_{tot} = S_b + S_c$ is the total gravitational entropy, 
  and a trough in $dM/dT_b$\footnote{ There is a minimum in $M(T_b )$ rather than a maximum due to the negative specific
  heat.}.  These extrema can be understood as a result of suppression of classical fluctuations in
$S_{tot}$ and $M$ at high and low temperatures \cite{Dinsmore:2019elr}. Since quantum fluctuations are 
always present in the system, particle production in SdS is a natural  way to extend the classical considerations,
and the high and low temperature limits are studied in detail here.
Since entropy is a property of the horizon geometry one needs to use particle states that are well behaved
 on the horizons. There have been several recent papers that anaylze closely related questions.
 The stress-energy tensor for the Unruh state in de Sitter has been computed \cite{Aalsma:2019rpt}
using Kruskal modes on the horizon,
and it would be interesting to see how those results extend to SdS.
A Schottky anomaly for AdS black holes has recently been studied in \cite{Johnson:2019vqf}\cite{Johnson:2019ayc}, 
and similar features
were computed for AdS black holes in two-dimensional dilaton gravity \cite{Grumiller:2014oha}. Another recent 
paper studies the Shottky anomaly in SdS and analyzes the system as a heat engine \cite{Johnson:2019ayc}.

Our results contrast with another recent calculation of the black hole and the cosmological  particle 
production in SdS  \cite{Bhattacharya:2018ltm}, where a particular 
set of states are chosen such that both spectra \textit{are} thermal at the temperatures $T_b$ and $T_c$. 
In this case, the particle states used are not well behaved on the horizons, so the calculations are done near the horizons.
Is there a consistency problem between the differing results? Consider the standard  Minkowski vacuum
 with particles defined by freely falling observers. Recall that  Rindler obervers, $i.e.$ ones who
  move with constant acceleration in this vacuum,
 detect a thermal flux of particles with temperature proportional to their acceleration.
 The temperature can be tuned by changing the acceleration.
So clearly the form observed for a particle spectrum depends 
on who does the measurement, so there is no contradiction. Nonetheless, it is reasonable to ask
how  these qualitatively different results are related.  To analyze this question, we introduce the notion of a spectral amplitude,
which is the complex quantity that leads to the (real) particle spectrum. 
We show in section
 (\ref{sectionexact}) that the spectral amplitude
for the Kruskal particles is equal to an integral over frequencies of the product of thermal amplitudes at temperatures
$T_b$ and $T_c$, weighted by an appropriate transmission coefficient.  Hence the amplitude for  produced particles 
crossing one horizon or the other can be thought of as composed of ``intermediate" thermal state interactions.

Particle production calculations are limited by the difficulty of solving the wave equation in a curved spacetime.
For Schwarzschild, Hawking argued that the geometric optics approximation can be used since particle production
is a high frequency process. However, SdS
has two scales which means that it is less clear what ``high" frequency means.  The wave propagation needs to be more
carefully studied to choose accurate frequency cut-offs. We find that introducing the intermediate states just referred to
is very useful in sorting out the cut-offs, as the transmission coefficients for the waves necessarily enter, and
these encode how modes behave as a function of  frequency.

The family of SdS spacetimes interpolates between
very small black holes, whose  geometry near the black hole horizon
 is like Schwarzschild, and large black holes with area that
approaches the area of the cosmological horizon.
Equipped with physical cut-offs
we are able to compute the black hole and cosmological particle spectra for this range of black hole areas.
The total rates of particles crossing each horizon is then found,  which requires some interesting considerations about the emission and absorption geometries. Small black holes with $T_b \gg T_c$
 share the black hole instability of Schwarzschild and emit particles in the expected  thermal spectrum, which is 
 compensated for by only a tiny amount of absorption of cosmological particles.
  In the large black hole limit both temperatures approach the 
 common value of zero in a quasi-equilibrium state and the spectra become equal, but particle production is exponentially
 suppressed. Hence these spacetimes illustrate a rich variety of complicated behavior.
 
This paper is organized as follows. In Section (\ref{setup}) we review relevant properties of SdS, choose the
particle states, and develop neede formulas to compute the particle spectra and total particle production rates.
 In Section (\ref{geomoptics}) the geometric optics approximation is used to evaluate the spectra in terms
 of an undetermined cut-off frequency. Exact formula
for the Bogoliubov amplitudes as the convolution of thermal amplitudes is derived in
Section (\ref{thermamps}) and physical frequency cut-offs are then inferred. The relevant density of states is derived in
Section (\ref{staphav}) and then the spectra and production rates are computed, with a focus on the small and large
black hole limits.
Section (\ref{conclusion}) presents our conclsions and open questions. Several details of the calculations are given
in the appendices.

\section{Setting up the Calculation}\label{setup}

We study a massless scalar field in SdS, and 
consider the causally connected region outside the black hole and inside the cosmological
 horizon as shown in Figure 1. Compared to an asymptotically flat black hole the boundaries at past and future null infinity 
 are replaced by the past and future
 cosmological horizons. Particles crossing the future cosmological horizon are interpeted as coming from the black hole,
 and those entering the black hole are interpeted as coming from the cosmological horizon. In this section
 we define the particle states and establish the formulae for  the particle spectra \ref{numberopsthree} 
 and the total rate
 of particle production (\ref{rate}). Some details are moved to Appendix (\ref{discrete}).

\subsection{Schwarzschild-deSitter basics}
The metric of Schwarzschild- deSitter spacetime in static patch coordinates $(t,r)$ is
\be\label{metric} 
ds^2=-f(r)dt^2+\frac{1}{f(r)}dr^2+r^2d\Omega ^2 
\ee
where
\be\label{fofr}
f(r)=1-\frac{2M}{r}-  \frac{r^2 }{ l_c^2} = \ - \frac{1}{ r_c^2} (r-r_c ) (r-r_b )(r +r_c +r_b )
\ee
where $r_c > r_b $ are the locations of the cosmological and black hole horizons respectively, and
 $ \frac{1}{3}\Lambda =1/l_c^2$. The two parametrizations are related by
 \be
 M = {r_br_c(r_b+r_c)\over 2 (r_b^2+r_c^2+r_br_c)},\qquad l_c^2 = r_b^2+r_c^2+r_br_c
\ee
The horizon surface gravities (or temperatures) are given by
 $2\pi T_h = \kappa_h =  | f' (r_h ) |/4\pi $, or
\begin{equation}\label{temperatures}
\kb ={(r_c-r_b)(2r_b+r_c)\over 2 l_c^2\, r_b},\qquad
\kc ={(r_c-r_b)(2r_c+r_b)\over 2 l_c^2\, r_c}
\end{equation}
Note that $\kappa_h$ is used to denote the $magnitude$ of the black hole and cosmological horizon
surface gravities, for $h$ respectively.
For the metric to describe a black hole horizon rather than a naked singularity, there is a maximum value to 
the mass
\be
M\leq M_{max} = {l_c\over 3\sqrt{3} }
\ee
which ensures that there are two positive roots to $f$. For fixed $l_c$, as $M$ increases from zero to its 
maximum, $A_b$ increases and $A_c$ decreases to the common value of $4\pi l_c^2 / 3$. The total entropy
$S= {1\over 4} (A_b + A_c )$ is largest for small black holes for which $T_b \rightarrow \infty$ and $T_c \rightarrow 1/l_c$. 
The entropy is smallest when the two horizons approach 
each other, in which case both temperatures  go to zero. Hence $S$ increases with increasing $T_b$ unlike
Schwarzchild black holes. On the other hand, the specific heat is still negative as with $\Lambda =0$.
The existence of a maximum size black hole is an important distinction from 
the $\Lambda \leq 0$ cases. For fixed $l_c$ 
 the full parameter space is explored by studying $0< r_b < l_c/ \sqrt{3} $, which will be referred to as
 small and large black holes.

The first laws in SdS  \cite{Dolan:2013ft} describe the influence of perturbations of one horizon on the other horizon
and on the mass, 
 \be\label{firstm}
\delta M = T_b \delta S_b - V_b {\delta \Lambda \over 8\pi} \ , \quad 
\delta M = - T_c \delta S_b - V_c {\delta \Lambda \over 8\pi}
 \ee
 Here $V_h$ is the thermodynamic volume associated with each horizon which in SdS
 has the simple form $V_h = {4\over 3} \pi r_h^3$. Taking the difference of the two equations
 gives a first law that only involves quanties in the causal diamond
\be\label{firstbetween}
 T_b \delta S_b +T_c \delta S_c = - V\delta \Lambda
 \ee
 where $V= {4\over 3} \pi  ( r_c^3 - r_b^3 ) $ is the thermodynamic volume between the horizons. Note that
  we are using the conventions that temperatures are positive and use $\kappa_h $ for the magnitude of
  the surface gravity at each horizon. This will avoid absolute value signs in expressions below. 
  Features of  black hole thermodynamics in deSitter, including  approaches to temperature, entropy,
  and conserved charges, are studied in 
  \cite{Dehghani:2002nt, Shankaranarayanan:2003ya,
Medrano:2007mg, Choudhury:2004ph, Myung:2007my, Urano:2009xn, ChangYoung:2010ps,
Kim:2014zta, Ishwarchandra:2014jca, Bhattacharya:2015mja, Li:2016zca, Hajian:2016kxx, Pappas:2017kam, Kanti:2017ubd, Robson:2019yzx}.
Features of black hole thermodynamics in an expanding universe are contained in
 \cite{Klemm:2015qpi, Gregory:2017sor, Gregory:2018ghc}. In addition to the classical gravitational perturbations described by the first laws (\ref{firstm}) and 
(\ref{firstbetween}), there are quantum
fluctuations in the matter fields, even if the classical values are zero. Our goal is to
 compute the quantum mechanical fluxes of particles crossing
  each horizon due to the presence of the other horizon.

\subsection{Choice of early and late time particle modes}\label{modechoices}

The first step is to choose modes functions associated with creation and
annhilation operators that define what is meant by a particle. We use the formalism for quantum field
theory in curved spacetime presented in
\cite{Birrell} and more specifically for black holes in
\cite{Traschen:1999zr}. Additional treatments are contained in  \cite{DeWitt:1975ys} \cite{Chao:1997em}.
The Penrose diagram for the portion of SdS bounded by the black hole and
cosmological horizons is shown in Figure 1.  Choosing the boundary
conditions on the mode functions is guided by the following physical considerations. In
Hawking's original calculation, with $\Lambda =0$, he considered gravitational collapse to form a black hole,
and so the past black horizon $\pbh$  was covered up by the collapsing star. This has the advantage that at early
times the geometry is close to flat, and the system starts in the usual Minkowski vacuum. 
At late times and far from the black hole, spacetime is close to Minkowski, so there is again a natural choice of 
positive frequency modes. Note that observers
near future null infinity who use the Killing time coordinate to define particles are freely falling.
It was later shown by Unruh \cite{Unruh:1976db} that the
particle production results are the same if
the calculation is done in the extended Schwarzschild spacetime by defining states on $\pbh$
 with respect to the geodesic null Kruskal coordinate $U_b$. This is a satisfying result not only 
because it is simpler to work in the extended spacetime than a collapsing star, but also because 
it illustrates that freely falling observers with coordinates defined by the geodesics are a choice that gives
physically sensible results. Unlike the far field, near the black hole geodesic time is not time along the Killing flow.

Consider the other limit, de Sitter spacetime with no black hole ($M=0$). There is a significant body of
research about choosing vacuum states in deSitter and inflationary cosmologies, 
see $e.g.$ \cite{Mottola:1984ar,Traschen:1986tn, Anderson:2000wx, Anderson:2001th, Collins:2003zv, Collins:2003mj,
 Leonard:2012si, Leonard:2012ex, Anderson:2013zia, Markkanen:2017abw, Das:2019aii, Aalsma:2019rpt}, and for a pedagogical treatment and further references see \cite{Birrell}.
 The quantum mechanical production of particles in inflation
 has been computed in multiple contexts and compared in detail to observations. In the cosmological
 setting, the picture is that a portion of an early FRW spacetime goes through a phase transition 
 which changes the effective equation of state to that of a cosmological constant, so
the spacetime is not de Sitter before a certain time or outside the 
 inflating bubble. The past cosmological horizon is covered up by an earlier FRW period.
  Similar to the fact that 
a collapsing star covers up $\pbh$, the entry to inflation covers up the past cosmological horizon $\pch$. 
Reference  \cite{Balbinot:1985mk} recently computed
properties of an Unruh state in two-dimensional de Sitter, using Kruskal coordinates to define modes 
on $\pch$ and $\fbh$.\footnote{In the two-dimensional case there is no scattering, and so
  boundary conditions on a portion of the past Cauchy surface can be 
 replaced with  boundary conditions on a portion of the future Cauchy surface, and generalizations of the 
 ``Unruh state" are often presented in this way. However in four dimensions there is scattering and these formulations
 are not the same.}.
 
Markovic and Unruh  \cite{Markovic:1991ua} extended the work of \cite{Unruh:1976db} for Schwarzchild to
two dimensional SdS, and computed
the two-point function and stress energy tensor, focusing on the observations of static observers, who are shown
to see a near thermal stress energy. They also demonstrated that there is agreement between
 using Kruskal boundary conditions on $\pch$ and $\fbh$ in the 
 extended two-dimensional SdS spacetime, and the results in a spacetime in which a collapsing star
  covers up $\pbh$.
   Reference \cite{Choudhury:2004ph} built upon the analysis
  of \cite{Markovic:1991ua} and examined the expectation value of the  stress-energy tensor for several choices 
  of SdS states in two dimensions. The response of particle detectors in black hole and in cosmological spacetimes
  was analyzed in \cite{Chakraborty:2019ltu}.  Our choice is the same as equation (26) of \cite{Choudhury:2004ph}
   in two dimensions, though in four dimensions  there are differences (see footnote \#2).
  Reference \cite{Choudhury:2004ph} finds that static particle detectors detect thermal fluxes.
 Hence the results of  \cite{Markovic:1991ua}, \cite{Choudhury:2004ph}, and
 \cite{Bhattacharya:2018ltm} make a consistent package, and report on thermal properties
 of several quantities measured by static observers 
  
In contrast,
as mentioned above, we define partilcles near each horizon with respect to Kruskal coordinates
of freely falling observers. The issue of static time $vs.$ geodesic time particles will be 
 returned to in Section (\ref{nonthermamps}), and connections made between our results to this earlier work. 
 Proceeding, we assume that the proper physics is captured 
 by doing the calculation in the extended
 SdS spacetime and defining positive frequency on $\pch$ with respect to the geodesic Kruskal coordinate $V_c$ and on
the $\pbh$ with respect to $U_b$, as was done in previous work of one of the authors \cite{Kastor:1993mj}.
Note that using these free-fall coordinates to define particle states on both horizons
 when $\Lambda >0$ 
 is a simple generalization of Unruh's analysis for Schwarzchild black holes \cite{Unruh:1976db}. The primary distinction
 from the other treatments is that we focus on produced Kruskal particles, rather than quantities measured by
 static, accelerating, observers.

Particles produced due to a horizon in one part of a spacetime are observed in another part of the spacetime.
In Schwazschild, particles interpeted as coming from the black hole 
 are detected at late times at future null infinity ${\cal I}^+$. In SdS  ${\cal I}^+$ is a spacelike surface
 to the future of $\fch$. We will restrict our attention to the causal diamond for SdS shown in Figure 1, in which
 the boundaries   ${\cal I}^\pm$ of an asymptotically flat spacetime
 are replaced by the cosmological horizons ${\cal H}^\pm_c$. Of course, in SdS the spacetime extends beyond $\fch$ and
it would be  interesting to compute the particle production observed by cosmological observers in the far field, 
 but in this paper we focus on the bath of particles near each horizon,
 due to the presence of the other horizon. 

 the modes are solutions to the wave equation
\be\label{waveeq}
g^{ab} \nabla_a \nabla_b f = g^{ab} \nabla_a \nabla_b p =0
\ee
and set of modes $f^h_\omega, f^{*h}_\omega $ and $p^h_\omega , f^{*h}_\omega$ 
forms an orthonormal basis in the conserved Klein-Gordon inner product,
\be\label{on}
(f^h_\omega , f^{h'}_\nu )   = \ \delta (\omega - \nu) \delta^{hh'}
\ee
where 
\be\label{innerprod}
(h, g) \equiv -i \int_\Sigma d^3 x \sqrt{ |\gamma |} n^a \left( h \partial _a g^* -g^* \partial_a h\right)
\ee 
Here $\Sigma$ is a Cauchy surface with unit normal $n^a$ and $\sqrt{ |\gamma |}$ is the volume 
element of the induced metric on $\Sigma$.

The null geodesic, or Kruskal, coordinates near each horizon are related to the static patch null coordinate
 $v=t+r^* $ and $u =t-r^* $ where $dr^* =dr/f$, by
\be\label{kruskals}
U_c=\frac{1}{\kappa_c}e^{\kappa_{c}u}\ ,  \quad V_c=-\frac{1}{\kappa_{c}}e^{-\kappa_{c}v} \ ,\quad
 U_{b}=-\frac{1}{\kappa_{b}}e^{-\kappa_{b}u} \ ,\quad V_{b}=\frac{1}{\kappa_{b}}e^{\kappa_{b}v}
\ee
One each horizon the appropriate Kruskal coordinate is equal to zero, see Figure 1.
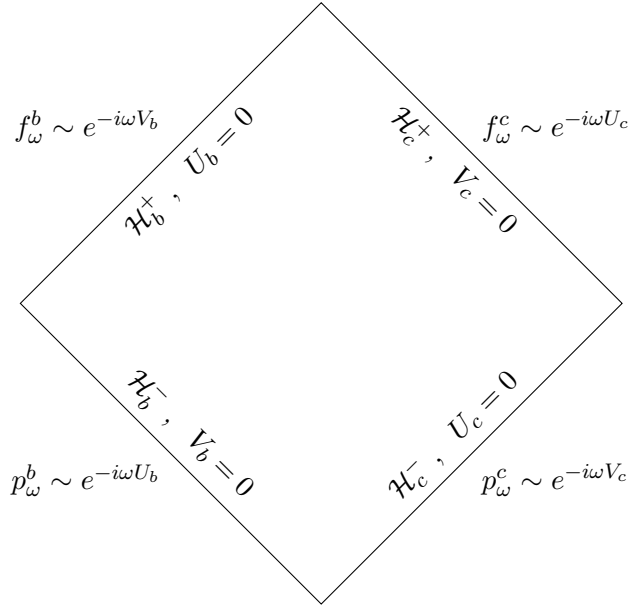
\begin{figure}[h]
\begin{tikzpicture}
\node (I)   at (0,0)   {};

\path  
  (I) +(90:4)  coordinate[label=90:$$]  (IItop)
       +(-90:4) coordinate[label=-90:$$] (IIbot)
       +(0:4)   coordinate                  (IIright)
       +(180:4) coordinate[label=180:$$] (IIleft)
       ;
\draw (IIleft) -- 
          node[midway, above left]    {$f_{\omega}^b \sim e^{-i\omega V_{b}}$}
          node[midway, below, sloped] {$ \fbh \ ,\ \ U_{b}=0$}
      (IItop) --
          node[midway, above right]    {$ f_{\omega}^c \sim e^{-i\omega U_{c}}$}
          node[midway, below, sloped] {$\fch \ ,\ \ V_{c}=0$}
      (IIright) -- 
          node[midway, above, sloped] {$\pch \ ,\ \ U_{c}=0$}
          node[midway, below right]    {$p_{\omega}^c\sim e^{-i\omega V_{c}}$}  
      (IIbot) --
          node[midway, above, sloped] {$\pbh \ ,\ \ V_{b}=0$}
          node[midway, below left]    {$p_{\omega}^b \sim e^{-i\omega U_{b}}$}    
      (IIleft) -- cycle;



\end{tikzpicture}

\caption{The "Causal Diamond", or static patch of SdS,
 bounded by the past and future black hole and cosmological  horizons.
Positive frequency particle modes are defined with respect to the appropriate Kruskal coordinate on each
portion of the boundary as indicated.}
\end{figure}

The union of the past black hole and cosmological horizons is a Cauchy surface 
for the diamond on which the boundary conditons for the early time modes.
Let $p_\omega^b $ denote a mode that
has positive frequency $\omega$ with respect to the affine coordinate $U_b$ on the past black hole horizon $b$ and
that vanishes on the past cosmological horizon,
\bea\label{modebcs}
p^b_{\omega lm}  & = & {Y_{lm} (\Omega ) \over \sqrt{4\omega } r_b } e^{-i\omega U_b }  \ \quad on\  \pbh 
\nonumber \\
&=& 0  \quad \ \quad \quad \quad on\  \pch \nonumber \\
\eea
where the $Y_{lm} (\Omega )$ are the spherical haromoincs on $S^2$. The modes are normalized in the Klein-Gordon innerproduct (\ref{innerprod}). One technical point is that, for example, the Kruskal coordinate $U_b$ ranges between
$-\infty < U_b \leq 0$ on $\pbh$. Under the Kruskal extension that surface is continued into the black hole region,
coverd by the other half-line $0\leq U_b < \infty$. The inner poduct for the $p^b_\omega$
 is taken along the entire range of the Kruskal coordinate\footnote{ We would like
 to thank Paul Anderson for useful conversations on this issue.}. However, except for the normalization, the 
 behavior of the modes in the extended region will not be relevant for our calculations, which examine
 what happens in the causal diamond and on its boundaries. Analogous comments pertain to the normalization 
 of the other sets of modes.

Similiarly, $p_\omega^c $ denotes a mode that
has positive frequency $\omega$ with respect to the affine coordinate $V_c$ on the past cosmological horizon $c$ and
that vanishes on the past black hole horizon. A second, late time Cauchy surface is the union of the future black hole and
cosmological horizons, and we denote the late time modes functions by $f$. The mode $f_\omega^c $ has positive
frequency $\omega$ with respect to $V_c$ on the future cosmological horizon and vanishes on the future 
black hole horizon, and conversely for $f_\omega^b $. The boundary conditionns are summarzed in Figure 1.

The field $\Phi$ can be decomposed in either the basis of the past or the future modes, 
 \begin{eqnarray}\label{phiexpand} 
\Phi&=& \Sigma_{lm} \int d\omega (a_{\omega lm}^b p_{\omega lm}^b
+a_{\omega  lm}^c p_{\omega lm }^c +h.c.)\nonumber\\
&=& \Sigma_{lm}\int d\omega (b_{\omega ,lm}^b f_{\omega lm}^b 
+b^c_{\omega lm }f^c_{\omega lm}+h.c.)\nonumber\\
\end{eqnarray}
Here $ a_{\omega ,lm}^{c\dagger} $ creates a particle at early times defined with respect to the Kruskal coordinate $U_c$
near the past cosmological horizon, 
and $ a_{\omega ,lm}^{b\dagger} $ creates a particle at early times defined with respect to the Kruskal coordinate $V_b$ near 
the past black hole (white hole) horizon. Likewise, $b_{\omega lm}^{b\dagger}$ and $b_{\omega lm}^{c\dagger}$ create
late time particles near the black hole and future cosmological horizons respectively.
Unless needed, we will suppress the
angular quantum numbers $l,m$.

\subsection{Formalism: black hole and cosmological particle spectra and total production rates }
Assume that the spacetime is intially in the early time vacuum defined by
\be\label{vacdef} 
a^b_{\omega} |0> =a^c_{\omega} |0> =0
\ee
Our goal is to compute the number of late time particles which are created by $b^{b \dagger}$  
near the future cosmological horizon,  which are interpeted by local observers as coming from
the black hole. Likewise, we will compute the late time particles created by  $b^{c\dagger}$ crossing the black hole horizon,
which are interpeted as being produced by the cosmological horizon. So, compared to an aymptotically flat black hole,
in this analysis the boundary at future (past) null infinity is replaced by the future (past) cosmological horizon.

Each late time mode can be expanded in terms of the early time basis and {\it vice-versa}. Let 
\be\label{expandmodes}
f^c_{\omega}=\int d{\omega'}\left[ \alpha^b_{\omega \omega'}p^b_{\omega'}+\ \beta^b_{\omega \omega'}p_{\omega'}^{b*}
+\ A^c_{\omega \omega'}p^c_{\omega'}+\ B^c_{\omega \omega'}p_{\omega'}^{c*} \right]
\ee
and similarly for the mode $f^b_{\omega}$ with the labels on the right hand side interchanged $( b\rightarrow c , \ 
c \rightarrow b )$. The mode-mixing Bogoliubov coefficients $\alpha_{\omega \omega'}$ and $\beta_{\omega \omega'}$  give the amplitude for scattering an in-modes with positive frequency $\omega$ to out-modes
with positive frequency $\omega^\prime$ and negative frequency $-\omega^\prime$ respectively, the latter 
leading to particle production. The $A^c_{\omega \omega'}$ and $B^c_{\omega \omega'}$ coefficients
are needed in the basis expansion and
are given by the overlap of a wave that enters the diamond through the 
past cosmological horizon, scatters off the geometry, and leaves through the future cosmological horizon.
This is a classical scattering process and does not contribute significantly to the the particle production, but
does enter into the normalization condition (\ref{norm}) below.

The Bogoliubov coefficients in (\ref{expandmodes}) are computed by using the Klein-Gordon inner product, 
\begin{eqnarray}\label{coeffs} 
\alpha^b_{\omega \omega'} & = & (f^c_{\omega}, p^b_{\omega'}) \quad 
\beta^b_{\omega \omega'}= -(f^c_{\omega}, p^{b*}_{\omega'})= -i \alpha_{\omega, -\omega'}\quad \nonumber\\
A^c_{\omega \omega'} & = & (f^c_{\omega} ,  p^c_{\omega' } )\quad
B^c_{\omega \omega'} =-  (f^c_\omega ,  p^{c*}_{\omega' }) \nonumber\\
\end{eqnarray}
The expressions for the coefficients $\alphac , \betac , A^b_{\omega \omega' }$ and $B^b_{\omega \omega' } $
are obtained by interchanging the labels $b$ and $c$.
Then the relation between the in and out particle operators follows from (\ref{expandmodes}) and (\ref{phiexpand}),
which give
\be\label{expandops} 
b^c_{\omega}=\int d{\omega'}\left[ \alpha_{\omega \omega'}^{b*} a^b_{\omega'}-
\beta_{\omega \omega'}^{b*} a_{\omega'}^{b\dagger} 
+ A_{\omega \omega'}^{c*} a^c_{\omega'}-
B_{\omega \omega'}^{b*} a _{\omega'}^{c\dagger} \right]
\ee
and similarly for the operator $b^b_{\omega}$, with the labels on the right hand side interchanged 
$( b\rightarrow c , \ c \rightarrow b )$. 

In equation (\ref{phiexpand}) the field $\Phi$ is expanded 
 in mode functions labeled by a continuous frequency parameter $\omega$,
which is convenient for calculations and we will be used here.
However, to get formulae for physically relevant quantities such as the number of particles
produced per unit volume, one needs to use properly normalized wave packets. Since many of the equations -- but not all-- repeat those already presented
 with discrete rather than continuous indices, we include the steps in Appendix (\ref{discrete}). The frequencies 
are integer-indexed as $\omega_j = j/R$ where $R$ is a length scale associated with the density of states, see
(\ref{density}). The key output is that 
the probablity of  horizon $h$ producing a late-time particle with  frequency $\omega_j$ which is observed
crossing horizon $h'$,  
 \be\label{numberopstwo}
N_{\omega_j } ^{ h } = \Sigma_k \  |\beta^h_{j k } |^2 \   
  \ , \ \quad h=b,c
\ee
becomes
\be\label{numberopsthree}
N_{\omega }^{ h } = \ {1\over R} \int d\omega' |\beta^h_{\omega \omega'  } |^2 
\ee
in the continuous basis, where $\sum_k = R\int d\omega' $ has been used.
Note the important dimensionful factor of $1/R$ that multiplies the integral using the continuous basis functions, and
that $N_\omega ^{ h } $ is dimensionless. 

The angular momentum eigenvalue $l$ has been mostly suppressed, but now we need to include the density of 
states. The total number of particles produced with frequency $\omega _j $
 is the product of  $N_{\omega_j}$ times the density of states $\rho (\omega _j ) = \Sigma_l (2l+1)$.
  In the continuous basis functions one has
 \be\label{density}
 \rho (\omega ) = \sum_l (2l+1)  (d\omega R ) 
 \ee
  The number of particles produced per unit time results from integrating (\ref{numberopsthree})
   over $\omega$ and dividing by $R$, which cancels the $R$ in the density of states,
  \be\label{rate}
  n_h =  \sum_l (2l+1) \int d\omega  N_\omega^h 
  \ee
This has dimensions $(length)^{-1}$ as it should.
The energy emitted per unit time $E^\phi_h$ in the  $\phi$-particles is gotten by including another factor of $\omega$ in the integrand of (\ref{rate}).

Lastly, one needs to fix $R$ that arises in the density of states, and appears in (\ref{numberopsthree}).
Reference \cite{Birrell} computes the
particle production due to a shell that collapses to a black hole, and  take $R$ to 
be the light travel time from the shell to a distant sphere where the particle flux is measured.
In analogy, we will take
$R$ to be the light travel distance between the two horizons. Studying null geodescis one
finds that $ R= (r_c - r_b)$, so for small black holes $R\simeq l_{cos}$. For large black holes
this distance is going to zero, and expressed in terms of the surface gravity one finds
 $R\simeq \kappa l_{cos}^2$, where we have set $ \kappa =\kappa_b \simeq
\kappa_c \rightarrow 0$.

\section{Particle production in the geometric optics limit}\label{geomoptics}

 Particle production results when a mode that is positive frequency with respect to the modes on one horizon
 propagates into a mixture of 
 positive and negative frequency modes on the other different horizon.
The  $\alphab$ mode mixing integral for production due to the black hole horizon is given by the inner product
 equation (\ref{coeffs}),
which we will evaluate on the past Cauchy surface. 
Since $p^b$ vanishes on $\pch$, this reduces to an integral over the
past black hole horizon. Hence we need the solutions for the future modes $f^c$ on $\pbh$.
Hawking argues \cite{Hawking:1974sw} that since the main effect comes from high frequency modes
 the wave equation can be solved in the geometric optics approximation, in which the wave $f^c$
propogates unchanged  on a surface of constant phase $U_b = constant$ to intersect $\pbh$. Explicitly,
using (\ref{kruskals}) to
trace a line of constant $U$ or $V$ in the causal diamond above, the Kruskal coordinates on the 
boundaries are related to each other by
\be\label{kruskalrel}
U_{c}=\frac{1}{\kappa_{c}}(-\kappa_{b}U_{b})^{-\frac{\kappa_{c}}{\kappa_{b}}}
 \quad and \quad V_{b}=\frac{1}{\kappa_{b}}(-\kappa_{c}V_{c})^{-\frac{\kappa_{b}}{\kappa_{c}}}
\ee
A mode $f^c_\omega $ goes like $ e^{-i\omega U_c }$ on $\fch$, so on $\pbh$ it goes like
$e^{ -i(\omega /kc ) (-\kb U_b )^{-\kc/ \kb } }$.
Substituting into the inner product (\ref{innerprod}) gives
\be\label{alphabgo}
 \alpha^b_{\omega \omega'}  =  {1\over 4\pi \sqrt{\omega\omega'}\kb } 
\int^{\infty}_0  dx e^{ -i  {\omega' \over \kb }x }e^{-i {\omega \over\kappa_c}x^{-\kappa_c /\kappa_b } }
\left(\omega'+\omega x^{-\frac{\kappa_T}{\kappa_b}}  \right) 
\ee
where $\kappa_T= \kappa_b +\kappa_c$. The integral for $\alpha^b_{\omega \omega'}$ only goes over 
$\pbh$ which is the half-line $-\infty < U_b \leq 0$, since no portion of the mode $f_\omega^b$ comes out of the black hole. 
It is straighforward to check that the integral for $\alphac$, for the particle production due to the cosmological horizon,
is given by interchanging $\kappa_c $ and $\kappa_b$ in (\ref{alphab}). 

It is interesting to compare $\alphab$ to the corresponding integral for an asymptotically flat black hole, which is given 
by setting $\epsilon =0$
in equation (\ref{alphasmall}), and  can be evaluated in terms of $\Gamma$-functions as given in
(\ref{base}). The Minkowski null coordinate used to define particles at future null infinity 
 is related by a single red shift to the Kruskal coordinates that define particles near the black hole horizon.
 In SdS the inner product is between two sets of Kruskal modes, which involves
a Kruskal-to-static coordinate redshift followed by a static-to-Kruskal  blueshift given in equation (\ref{kruskals}). 
The sequence of red and blue shifts gives a more complicated integrand, and produces a spectrum that is not thermal.

For generic values of the surface gravities these integrals can be evaluated using the method of
 stationary phase,
\begin{eqnarray}\label{stapha} 
I(k )&=&\int e^{i  \phi(t) /k } f(t)\nonumber\\
&\overset{k \to 0}{\simeq}&e^{i \phi(t_s)/ k }f(t_s)e^{Sign(\phi''(t_s))\frac{i\pi}{4}}\sqrt{\frac{2\pi k }{|\phi''(t_s)|}}
\end{eqnarray}
where $t_s$ is the point of stationary phase $\phi ' (t_s ) =0$ which is assumed to be in the range of integration.
 Applying (\ref{stapha}) to equation (\ref{alphabgo}) with $\phi (x) = -(\omega' x \kc/ \kb +x^{-\lambda}\omega )$
  and $x_0 = (\omega / \omega' )^{\kb / \kt } $ gives
\begin{equation}\label{alphab}
\alpha^b_{\omega \omega'} \simeq 
 -{   \sqrt{2} e^{-\frac{i\pi}{4}} \over \sqrt{ \pi  \kappa_T } (\omega^{\kc} \omega^{\prime\kb })^{1/2\kt } } 
 \exp [ -i {\kappa_T \over \kappa_b \kappa_c }
 (\omega^{\prime\kappa_c}  \omega^{\kappa_b}  )^{1/\kappa_T} ]
\end{equation}
As summarized in equation (\ref{coeffs}) one then analytically continues in $\omega'$ to get
 $\beta^b_{\omega \omega'} = -i \alpha^b_{\omega, - \omega'} $. Viewed as a complex function of $\omega'$,
  $\alpha_{\omega \omega'} $  is analytic in the lower half $\omega'$-plane, since it is the Fourier transform of 
the function (\ref{alphab}), which vanishes in the lower half $x$-plane. Hence 
when analytically continuing the branch cut that arises must be put in the upper half $\omega'$ plane, and
$\omegap \rightarrow -\omegap = e^{-i\pi} | \omegap|$, giving
\begin{equation}\label{betabgo}
\beta^b_{\omega \omega'} \simeq 
-i {   \sqrt{2} e^{-\frac{i\pi}{4}}  e^{i\pi \kb /2 \kt} \over \sqrt{ \pi  \kappa_T } (\omega^{\kc} \omega^{\prime\kb })^{1/2\kt } } 
 \exp [- {\kappa_T \over \kappa_b \kappa_c }
 (\omega^{\prime\kappa_c}  \omega^{\kappa_b}  )^{1/\kappa_T} (i \cos \pi {\kappa_c \over \kappa_T} 
 +\sin \pi {\kappa_c \over \kappa_T} ) ]
\end{equation}
This Bolgoliubov coefficient gives the mixing of negaitive frequency modes  the black hole horizon with
positive frequency cosmological modes, that is, a cosmological Kruskal observer near $\fch$ interpets the resulting
flux as particles coming from the black hole.
 To get the reverse situation of particles detected near the black hole produced from the 
cosmological horizon, one can check that $\kb$ and $\kc$ are simply interchanged in (\ref{alphab}), 
\be\label{alphacgen}
\alphac (\kb , \kc ) =\alphab (\kc , \kb)
\ee

Interstingly, after using the stationary phase approximation, the integral for the number of produced particles
(\ref{numberopsthree}) can be done easily, giving
%
\be\label{Nbgenone}
N^{b}_\omega  =  {2\over \pi \omega R}{\K \over \kc }
\exp [-{1\over {\cal K} }(\omega^{\kb}  \omega_0^{\kc}  )^{1/\kt}   ]   \ \  \  , \ 
\ee
where
\begin{equation}\label{Kdef}
    \K =   { \kc\kb \over 2\kt  sin(\pi \kc /\kt ) } \rightarrow
    \begin{cases}
   \kb / 2\pi , &  \kb \gg \kc \\
  \kappa /4  & \kappa= \kb\rightarrow \kc \rightarrow 0 
       \end{cases}
  \end{equation}
  Hence $\K$ is equal to the black hole temperature in the small black hole limit, and proportional to the common
temperature for large black holes.
The spectrum for the cosmological particles $N^c_\omega $ is obtained by interchanging 
$\kb$ and $\kc$. Note that the quantity $\K$ is invariant under interchanging $\kb$ and $\kc$.

This is as far as we can go without specifying the cut-off frequency in the integrral over $\omega'$ to get $N_\omega$.
 In Schwarzchild there is only one scale, so 
neccessarily $\omega_0 \propto 1/r_b$. However SdS has two scales, so a better understanding of the
process is needed, which we now turn to.

\section{Exact treatment: thermal amplitudes, non-thermal amplitudes, and physical cut-offs}\label{sectionexact}

Consider an early time mode $p^b_\omega$ (or $p^c_\omega$) that has 
boundary conditions set on the past horizons. 
The inner product for a Bogoliubov coeffient is an integral over a Cauchy surface. So to compute the mode mixing
with a late time mode one needs to know the solution for $p^b_\omega$ on the future horizons, and this
is where the technical challenges arise. Following Hawking \cite{Hawking:1974sw}, the wave equation was
solved in the geometric optics limit in the previous section.  We next show that the geometric optics approximation
plus judicious choices for the low frequency cut-offs and for the limits on the sum over angular momentum
quantum number $l$ in the integrals for $N^h_\omega$  and $n_h$ give accurate results for the particle production.
If one could solve the wave equation
then the solutions for the modes would incorporate 
 such restrictions, as will be clear in results below. But finding solutions is a hard problem so approximation techniques must be used.  So the goal of this section is to
 derive an exact expression for the Bogoliubov coefficients that includes the effects of
scattering of the modes. 
This also leads to interesting physics.
 
\subsection{Building blocks: thermal amplitudes }\label{thermamps}

In the particle production calculation the real valued spectrum  $N_\omega$ results
from squaring the complex quantity $\beta_{\omega\omega'}$ 
and summing over the contributions from all states $\omega'$. In analogy
with the relation between a wave function and the expectation value of an operator, 
we introduce the notion of $\beta_{\nu\nu'}$ as the ``amplitude for a spectrum", that is, 
the complex function whose integrated norm gives the spectrum,
\be\label{ampdef}
N_\nu  = {1\over R} \int d\nu' |\beta_{\nu \nu' } | ^2
\ee
The amplitude depends on the background geometry and on the choices of the two bases of past and future modes
that are being compared. An important reference case is the asymptotically flat
Schwarzchild black hole \cite{Hawking:1974sw}. Then the amplitude is the inner product between black hole $Kruskal$ 
modes with outgoing null time coordinate $U$, 
and asymptotically $Minkowski$ plane waves near future null infinity $\cal{I}^+$
with outgoing null time coordinate $u$. These
dependences will be indicated as $\beta_{\nu\nu' }^{AF} (K_b , M ;\kb )$ where $K_b$ refers to Kruskal coordinates
near the black hole horizon and $M$ refers to freely falling Minkowski coordinates.
In the geometric optics approximation Hawking showed that the amplitude has a nice
expression in terms of Gamma-functions,
\bea\label{base}
 \beta_{\nu \nu' }^{AF} (K_b , M;\kb ) & =& e^{-\pi \nu /\kb } \alpha_{\nu \nu' }^{AF} (K_b ,M;\kb ) \ , \\ \nonumber
where\quad \alpha_{\nu \nu' }^{AF} (K_b , M ;\kb ) &=& 
  -{ 1 \over \pi \kb\sqrt{\nu\nu' } } \left( {i \nu'  \over \kb } \right)^{-i\nu / \kb}  \Gamma ( 1+{ i\nu\over \kb} )  \\ \nonumber
\eea
 This basic ``building block" for
Schwarzchild can be thought of as a $thermal$ spectral amplitude, in the sense that
 the resulting spectrum is, famously, thermal,
\be\label{specflat}
N_\nu^{AF}(K_b , M; \kb) = {1\over R} \int d\nu' |\beta_{\nu\nu'}^{AF} (K_b ,M; \kb) | ^2 =
 {\gamma_\nu^{AF}  \over e^{\nu/ 2\pi \kb} -1 }
\ee
where $\gamma_\nu^{Sch}$ is the classical scattering amplitude in the Schwarzchild geometry and
$\kb =1/2 r_b $ is the black hole surface gravity in Schwarzchild. 

A second relevant example studied recently \cite{Bhattacharya:2018ltm} 
 is the SdS black hole with the choice of modes chosen to
closely parallel the Schwarzchild calculation. The past modes on $\pbh$ are defined with respect to Kruskal time
$U_b$, but unlike our calculation,
 the future modes are defined with respect to the static patch null coordinate $u$ on a 
null surface near the cosmological horizon. It is found \cite{Bhattacharya:2018ltm} that
\be\label{equalamps}
\beta_{\nu\nu'}^{b, SdS} (K_b , M; \kb ) = \beta_{\nu\nu'}^{AF} (K_b , M; \kb)
\ee
One should note that the two surface gravities $\kb$ on each side of (\ref{equalamps}) do not have the same expressions
in terms of $r_b$ and $l$, since $\Lambda =0$ on the right hand side. 
It follows that the black hole spectrum for Kruskal to static modes is thermal \cite{Bhattacharya:2018ltm},
\be\label{specsdsu}
N_\nu^{b,SdS}(K_b, M; \kb ) = 
{1\over R} \int d\nu' |\beta_{\nu\nu'}^{b,SdS} (K_b , M; \kb ) | ^2 = \ {\gamma_\nu^{SdS}  \over e^{\nu/ 2\pi \kb} -1 }
\ee
where $\gamma_\nu^{SdS}$ is now the classical scattering amplitude in the SdS geometry.
Hence when the static coordinate plane waves are chosen for the future modes, 
 the difference in the amplitudes between  SdS
and Schwarzchild only appears in  the different greybody factors, for the same value of $\kb$. 
Reference \cite{Bhattacharya:2018ltm} also derives
an analogous expression for the cosmological particle production in SdS, choosing the past 
modes to be defined with respect to the ingoing Kruskal coordinate $V_c$ on $\pch$ and the future modes
defined with respect to the static patch
 ingoing null coordinate $v=t + r_* $ near $\fbh$. The resulting Bogoliubov coefficient has the 
 same functional form as that for asymptotically flat black hole emission with $\kb$ replaced by $\kc$, 
 \be\label{betacv}
\beta_{\nu\nu'}^{c,SdS} (K_c , M ;\kc ) =\beta_{\nu\nu'}^{AF} (K_b , M; \kc )
\ee
and so again leads to a thermal spectrum for the cosmological particle production at $T_c=\kc /(2\pi)$.

\subsection{ SdS emission as a convolution of thermal amplitudes}\label{nonthermamps}

We next derive an exact formula for the amplitude in SdS defined with respect to Kruskal modes on the past $and$
future horizons, which will turn out to be an integral over thermal amplitudes
 $\beta_{\nu\nu'}^{b, SdS} (K_b ,M ; \kb )$ and $\beta_{\nu\nu'}^{c,SdS} (K_c ,M; \kc )$ weighted by the transmission 
 coefficients for wave propagation in SdS. We will find that in general $N^b_{\omega}$ and $N^c_{\omega}$ are not thermal,
 but through this formula can be viewed as an integral over a continuous set of pre-thermal interactions.
The analysis takes advantage of  the fact that the wave equation is separable in the static patch $(t,r_*)$ 
coordinates\footnote{We thank Paul Anderson for pointing out this approach}. While
analytic solutions are not known in the $(t,r_*)$ coordinates either one replaces the problem of solving a
 partial differential equation with solving a  standard one-dimensional scattering problem, whose solutions 
can be written in terms of transmission $T_{\nu l}$ and scattering $S_{\nu l}$ coefficients that can be evaluated by various approximation techniques.
Including these coefficients in the calculation 
yields an exact expression for $\betab$ that has an interesting and useful form in terms of 
simpler $thermal$ amplitudes.

To proceed, we make use of a
third set of mode functions $k_\omega$ as ``intermediate states". In the static patch coordinates of 
the metric (\ref{metric}),
solutions to the wave equation $\Box k_\omega =0$ are separable as
\be\label{separate} 
k_{\nu l}(r_*,t)=\frac{\psi_{\nu l} (r_*)e^{-i\nu t}}{ r \sqrt{2\pi\nu}} Y_{lm} 
\ee
where $dr_* =dr/f$. The black hole horizon is at $r_*\rightarrow -\infty$ and the cosmological horizon at 
$r_*\rightarrow \infty$. $k_\nu$ is a solution to the wave equation if
$\psi_\nu$ is a solution to the ordinary differential equation
\be\label{waveone}
\partial^2_{r_*} \psi_\nu +\left( \omega^2 -V (r_* ) \right) \psi_\nu =0
\ee
where the potential is
\be\label{potential}
V= V_0 + V_l \ = \   {f\over r} { df\over dr} - {l ( l+1) \over r^2 }
\ee
 The $l$-independent part of the potential goes to zero exponentially fast in the tortoise coordinate near each horizon,
\be\label{expfast}
V_0\rightarrow  \kappa_b^2 e^{\kappa_b r_*}  \ , \quad near\ \ {\cal H}_b \ ; \quad \quad
V_0\rightarrow  - \kappa_c^2 e^{-\kappa_h r_*} \ , \quad near \ \ {\cal H}_c
\ee
 $V_0$ also is equal to zero at $r_0^3 = l^2M={1\over 2} r_b r_c (r_c + r_b )$ where  $df/dr =0$, so
   $V_0$ is positive for $r_b < r <r_0 $ where $f^\prime $ is positive, and
is negative for $r_0 > r > r_c $. This reflects the competition between the attraction of the black hole and the cosmological
expansion. 

\begin{figure}
  \includegraphics[width=11cm]{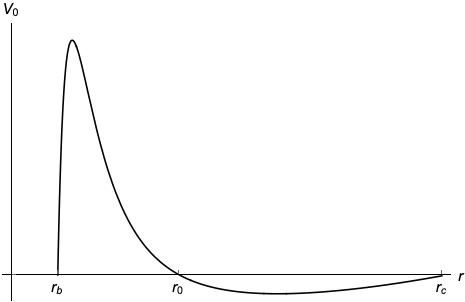}
  \caption{Plot of the potential $V_0$. There is a potential barrier and a potential well in Schwarzchild de Sitter space-time.  }
\end{figure}

As before, we suppress the angular momentum index unless needed for clarity.
Consider solutions for $k_\nu$ that  are plane waves in the
outgoing null coordinate $u=t-r_* $ near $\pbh$,  and vanish on $\pch$. Part of the wave then propagates across the
  potential barrier  and leaves through $\fch$ with amplitude $T_{\nu l}$, and a portion
scatters back into $\fbh$ with amplitude $S_{\nu l}$, so
\begin{equation}\label{kbc}
    k_{\nu l}\rightarrow
    \begin{cases}
   0 , &  \pch\\
      \frac{ 1 }{\sqrt{2\pi\nu}r_b}e^{-i\nu u}Y_{lm}  &  \pbh\\
   \frac{ 1}{\sqrt{2\pi\nu}r_c} T_{\nu l} e^{-i\nu u}Y_{lm}   &   \fch\\
      \frac{ 1  }{\sqrt{2\pi\nu}r_b} S_{\nu l} e^{-i\nu v}Y_{lm}  &  \fbh
    \end{cases}
  \end{equation}
 where $ |T_\nu |^2 +|S_\nu |^2  = 1$. 
At high frequencies the wave is well over the barrier and 
$T_\nu \rightarrow 1 , \ S_\nu \rightarrow 0$. This is regime in which geometric optics is applicable.
At low frequencies most of the wave
is scattered back by the potential and $T_\nu \rightarrow 0  ,\ S_\nu \rightarrow 1$.

Again, we start with the expansion of the future modes $f^c_\omega$ in terms of the past modes $p^b_\omega$
given in equations (\ref{expandmodes}) and (\ref{coeffs}). The new ingredient is to also
expand the $k_\nu$ in terms of the $p^b_\omega$, and the $f^c_\omega$  in terms of the $k_\nu$,
\bea\label{kexp}
k_{\nu} & =& \int_0^\infty d{\nu'}\left[ J_{\nu \nu'}p^b_{\nu'}+\ L_{\nu \nu'}p_{\nu'}^{b*} \right] \\ \nonumber
f^c_{\omega} & =& \int_0^\infty d{\nu'}\left[ M_{\omega \nu'} k_{\nu'}+\ N_{\omega \nu'}k_{\nu'}^* \right] \\ \nonumber
\eea
hence
\be\label{fcexpandtwo}
 f^c_\omega   =  \int_0^\infty d{\nu'} d\nu \left[ p^b_{\nu'} \left( M_{\omega \nu} J_{\nu \nu'} 
+ N_{\omega \nu} L^*_{\nu \nu'}\right)
+ p_{\nu'}^{b*} \left(  M_{\omega \nu} L_{\nu \nu'} + N_{\omega \nu} J^*_{\nu \nu'}\right)  \right] 
\ee
which implies via  (\ref{expandmodes}) that
\be\label{alphabmore}
\alphab =  \int_0^\infty d{\nu} ( M_{\omega \nu} J_{\nu \omega'} + N_{\omega \nu} L^*_{\nu \omega'} )
\ee
The analogue of the terms with the coefficients $A^c_{\omega \omega'}$
and $B^c_{\omega \omega'}$ which appear in (\ref{expandmodes}) have been dropped
since as mentioned earlier they make subdominant contributions to particle production.
 The new expansion coefficients are given by the inner products
\be\label{coeffstwo}
J_{\nu \nu'} =(k_\nu \ , p^b_{\nu'} ) \ , \quad M_{\omega \nu'} =( f^c_\omega \ , k_{\nu'} )\ ,\quad and\quad
L_{\nu \nu'} =-iJ_{\nu , -\nu'} \ ,\quad N_{\omega \nu'}= -i M_{\omega , - \nu'}
\ee
We evaluate $J_{\nu \nu'}$ on $\pbh$, 
\bea\label{jnn}
J_{\nu \omega' } &=& {1 \over 2 \sqrt{2} \pi\kb \sqrt{\nu\omega'} } \int_0^\infty dx e^{i \nu \ln x /\kb }
e^{-i\omega' x /\kb } \left(\omega' + {\nu \over x } \right) \\ \nonumber
&=& {-i\over\sqrt{2}  \pi \kb \sqrt{\nu\omega'} }
 \left( { i\omega' \over \kb }\right)^{-i\nu/ \kb } \Gamma (1+i {\nu \over \kb } ) \\ \nonumber
 & =&{ i \over \sqrt{2}} \alpha_{\nu\omega'}^{AF} (K_b ,M; \kb )
\eea
and evaluation of $M_{\omega \nu'}$ on $\fch$ gives
\bea\label{mnn}
 M_{\omega \nu} &=& {-iT^*_{\nu}  \over \sqrt{2} \pi \kc \sqrt{\omega\nu} }
 \left( { i\omega \over \kc }\right)^{-i\nu/ \kc } \Gamma (1+i {\nu \over \kc } ) \\ \nonumber
 &=& { i\over \sqrt{2} } T^*_{\nu}\  \alpha_{ \nu\omega}^{AF} (K_b ,M; \kc )
\eea
The transmission coefficient $T^*_{\nu}$ has entered through the future boundary behavior of $k_\nu$. 
%
Combining (\ref{alphabmore}), (\ref{jnn}), and (\ref{mnn}) gives  
\bea\label{alphabexact}
\alphab (K_b , K_c ; \kb , \kc ) & = & \int_{-\infty}^{\infty} d\nu \  T_\nu^*   \alpha^{AF}_{\nu\omega}(K_c, M;\kc )
\alpha^{AF}_{\nu \omega' } (K_b, M; \kb ) \\ \nonumber
 = &- & {1\over 2 \pi^2  \kb\kc \sqrt{\omega\omega'}}  \int d\nu {T_\nu^* \over \nu }\ 
\Gamma (1+ i{\nu \over \kb } ) \left(  i\omega' \over   \kb \right)^{- i\nu/\kb } 
 \Gamma (1+ i{\nu \over \kc } ) \left(i\omega   \over \kc \right)^{- i\nu/\kc }  \\ \nonumber
\eea
where the  two terms in (\ref{alphabmore}) have been combined by extending
range of integration over the entire $\nu$ axis and
 using $T_\nu^* = T_{-\nu} $. Analytically continuing in $\omega'$ as discussed earlier,
  $\omega' \rightarrow |\omega' | e^{-i\pi}$,
and substituting the relations (\ref{base}) and (\ref{equalamps}) gives the desired expression for $\betab$ 
\be\label{betabexact}
\betab (K_b , K_c ; \kb , \kc )  =  \int_{-\infty}^{\infty} d\nu T_\nu^*   
\alpha^{AF}_{\nu\omega}(K_c, M;\kc ) \beta^{AF}_{\nu \omega' } (K_b, M; \kb )
\ee
An analogous expression holds for the amplitude for the cosmological particle production.

What can be learned from the result (\ref{betabexact})?
At the beginning of this paper we argued that to study particle production at the horizons, particles must be defined
with respect to modes that are well behaved on the horizons. 
In general, the resulting spectra will not thermal, see
(\ref{Nbgenone}) (or (\ref{Nbgenagain}) and (\ref{Ncgenone})). However, the relation (\ref{betabexact}) shows that
the $amplitude$ for the spectrum can be written as a convolution of $thermal\  amplitudes$. To get an observable
one must take the complex modulus of $\betab$, which does not just depend on the squares of the thermal 
amplitudes, but depends on interference effects in the integral (\ref{betabexact}), which is expected
in a quantum process. What is perhaps less expected is that the quantum interactions
between the horizons can be summarized in such a simple form.

\section{ Spectra, density of states, and production rates}\label{staphav}
While (\ref{alphabexact}) and (\ref{betabexact}) provide physical insight, there
does not seem to be a route for evaluating the 
integrals exactly. So we will derive an approximate expression for (\ref{alphabexact}) again using 
the method of stationary phase (\ref{stapha}). The details are in Appendix (\ref{staphaapp}), where it is shown that after
three applications of stationary phase the triple integral for $\alphab$ reduces to the geometric optics result (\ref{alphab}) 
except that there is now a factor of the transmission coefficient  multiplying the expression. So the result for 
 $\betab$ is also the same as the geometric optics result (\ref{betabgo}) mutiplied by $T_\nu$,
  \begin{equation}\label{betabex}
| \beta^b_{\omega l \omega' l'} |^2 \simeq 
{2 |T_{ \sigma l'} |^2  \over \pi \kt (\omega^{\kc} \omega^{\prime \kb} )^{1/\kt } } 
exp[ -{ 2  \over \K}   (\omega^{\prime\kc} \omega^{\kb })^{1/\kt }  ]\ \delta_{ll'}
\ee
 where $ \sigma$ is the complex frequency
  \be\label{wstartwo}
   \sigma   = (\omega^{\prime \kc} \omega^{\kb} )^{1/\kt } e^{-i \pi\kc/ \kt}
   \ee
and $ \K =    \kc\kb [2\kt  sin(\pi \kc /\kt ) ]^{-1} $, see (\ref{Kdef}) for the limiting behaviors of $\K$.
The complex phase factor in $\sigma$ results from the analytic continuation in  $\omega' $ to obtain $\betab$ that leads to
the decaying exponential in $|\betab |^2$.

The integral over $\omega'$ gives 
\be\label{nbwitht}
N^{b}_{\omega l }   = \delta_{ll'} \  {2\over \pi \omega R}{\K \over \kc }  
 \int_0^{\infty} dx 
\ |T_{x' l'} |^2  e^{-x} \   \  ,  \quad\quad \quad  x' = \sigma (x) =\ x \K  e^{-i \pi\kc/ \kt}
\ee
We have restored the $l,l'$ indices to remind the reader that there is dependence on the angular momentum, as this will be important in the next section.
The spectrum for cosmological particles $N^{c}_\omega $   is obtained by interchanging 
$\kb$ and $\kc$. 

Equations (\ref{betabex}) and (\ref{nbwitht}) are equal to the corresponding quantities derived using
geometric optics times factors that depend on the transmission coefficient. 

The (simpler) geometric optics method gave
the result (\ref{Nbgenone}) for $N_\omega^b$, which involves a low frequency cut-off $\omega_0$.
Comparing (\ref{Nbgenone}) to (\ref{nbwitht}) implies that $\omega_0$
should be chosen equal to the frequency where the transmission coefficient changes from zero to one.
More precisely, we have shown is that the geometric optics method
plus choosing the cut-off to be in the transition regime of the transmission coefficient $T_{\omega l}$
gives a good approximation for the Bogoliubov coefficients. This is one of our main results.

 \subsection{Relevant density of states and processed formulae}
 
We now turn to the issue of finding the transition frequency for the transmission coefficient as just discussed.
We also identify additional conditions on $\omega$ and $l$ that must be incorporated before integrating over the product of
the density of states times $N_\omega^h$ to get $n_h$.

Even without detailed knowledge of the behavior of the transmission coefficient $T_{\omega' l}$,  
 equation (\ref{nbgen}) contains useful information. We know that
$ |T_{\omega'}l |^2 $ goes to one when the wave is well over the potential $V$ 
and goes to zero when $\omega' $ is sufficiently small compared to the maximum of $V$. The potential (\ref{potential}) has a 
 portion $V_0$ that is independent of the angular momentum of the field, plus an angular momentum barrier
 $V_l = l(l+1) / r^2$. A tractable approximation is that a
  mode with frequency $\omega'$ is over the barrier when its frequency squared is greater than the height of each
  portion of the potential,$i.e.$,  when  $\omega'> \omega_0$, where
  \be\label{lcutoff}
  \omega_0^2  \simeq Max \left( max (V_0 ) \ ,  {l^2 \over r^2_m} \right)
\ee
Here $r_m$ is the location of the maximum of $V$.  
We will refer to $\omega_0$ as the cut-off frequency, and approximate the spectrum   
 by replacing $T_{\omega' l}$ with a step-function at the cutoff. Then the black hole
 spectrum (\ref{nbwitht}) reduces to the expression derived using geometric optics  (\ref{Nbgenone}), with the 
important addition that now the cut-off is specified in (\ref{lcutoff})
 
For Schwarzchild the maximum of $V_0$ is at  $r_m = 4 r_b /3$, and 
for large black holes necessarily $r_m \simeq r_b$, so we will simply approximate  $r_m \simeq r_b $.
For small black holes the geometry near the black hole is close to Schwarzchild, and one finds that
 $ max (V_0 ) \simeq \kb^2 $. This is simply the physical result that the wave is over 
the barrier if its frequency is greater than the black hole temperature. On the other hand, for large black holes when
 $\kb\rightarrow \kc \rightarrow 0$,  one finds that the maximum of the potential goes to zero like $max (V_0 ) \simeq 
l_c^{1/2} \kappa^{3/2}$. Hence,
\be\label{cutoffsthree}
  \omega_0 \simeq {\kb \over 2\pi } \ \  ,\  small\ bhs \quad\quad and \quad\quad
   \omega_0 \simeq l_c^{1/4} \kappa^{3/4} \ \ , \ large\ bhs
  \ee
These considerations apply equally well to production of particles by the black hole or by the cosmological horizon, since
$V$ is fixed for a given geometry.

To get the total rate of particle production $n_h$ in (\ref{rate}) from $N_{\omega l}$ one sums over $l$ and integrates over
 $\omega$. There are constraints on the range of both $l$ and $\omega$, as follows.
 Start by writing equation (\ref{alphabexact}) for $\alphab$ as
\be\label{one}
\alphab =F( \omega , \omega' ; \kb , \kc )
 \ee
 Since  $\alphac$ results from interchanging the labels $b$ and $c$ one has
 \be\label{two}
\alphac =F( \omega , \omega' ; \kc , \kb )
 \ee
Inspection of (\ref{alphabexact}) shows that $F$ is symmetric under interchange of $\omega , \omega'$ and
$\kb , \kc$,
\be\label{three}
F( \omega , \omega' ; \kb , \kc ) =F( \omega' , \omega ; \kc , \kb )
\ee
which translates to
\be\label{four}
\alpha^b_{\omega \omega'} = \alpha^c_{\omega' \omega}
\ee
Since we have  established that the functions $\alpha^b_{\nu \mu}$ and $\alpha^c_{\nu \mu}$ vanish 
when the second frequency index is less than $\omega_0$, equation (\ref{four}) implies that there is also
a cut-off at $\omega_0$ in the first index of these functions.

Next, before summing over the angular momenta in the density of states the condition on $l$ in (\ref{lcutoff}) must be
incorporated. For fixed $\omega$, the condition implies that $l<\omega r_b$, hence
  \be\label{denistysum}
\int_{\omega_0} d\omega \  \sum_{l=0}^{ [r_b \omega ]} \ (2l+1) N^b_\omega \simeq 
r_b^2 \int_{\omega_0} d\omega \omega^2 N^b_\omega
\ee
It is satisfying that doing the sum over $l$ brings in the familiar factor of $\omega^2 $ times a $length^2$ 
in the density of states. 

In addition to the condition that the wave is over the barrier, there are features of the
absorption and scattering geometry that need to be taken into account.
For an oscillation to be detected as a wave by measurements inside the causal diamond its wavelength must be 
less than the propagation distance, that is, $\omega > 1 /( r_c -r_b )$. For small black holes this is not an
additional restriction compared to (\ref{lcutoff}), but for large black holes we have 
\be\label{omegafit} 
\omega >\omega_1 = {1 \over r_c -r_b } \simeq {1\over l_c^2 \kappa}  \ , \quad large\ bhs
\ee
which is a much higher cut-off than $\omega_0  = \sqrt{ max(V)}  \simeq l_c^{1/4} \kappa^{3/4}$.
 Since the condition (\ref{omegafit}) is about
a classical detection process it applies to the integration over frequencies of the spectrum to get $n_h$,
but not to the first integration over the frequencies of the amplitudes. It applies to both $n_c$ and $n_b$ since 
emission and absorption processes become symmetrical between the two horizons for large black hokes.

However not everything is symmetric when the black hole is small. 
Any wave propagating outwards from the black hole will
cross the cosmological horizon, but a  wave propagating inwards from the cosmological horizon needs to have a wavelength
$\lambda  < r_b$, or it simply sloshes over the black hole and propagates out through the 
cosmological horizon. A small length $\lambda$ corresponds to a large angular momentum
number $l$ as $\lambda  \simeq r_c /l $, so for a mode to be absorbed by the black hole 
\be\label{highl}
 l> {r_c \over r_b }
 \ee
 This is an important constraint for small black holes with $r_c /r_b \gg 1$. 
For larger black holes it is not a significant restriction, which corresponds to the fact that the black hole is 
indeed in the way of most waves.  
Combining (\ref{highl}) and (\ref{lcutoff}) gives the following allowed range of $l$ and more stringent cutoff $\omega_1$
for a small black hole to absorb cosmological particles,
 \be\label{lcutoffcos}
r_b \omega >\  l \  > {r_c \over r_b}= {1\over \epsilon} \gg 1 \ ,
 \quad\quad and \ \   \omega_1 = \kb {r_c \over r_b} ={\kb \over \epsilon} \gg \kb \ , \quad
(small\  bh) 
\ee 
where small black holes are defined by $\epsilon = r_b / r_c \ll 1$.

Let us close by summarizing the sum over contributing states. We use the notation $\omega_0  $ for the cut-off
derived from the condition that the wave is over $V_l$, which is the lower limit for the integration over $\omega'$
to get $N_\omega$. We use $\omega_1$ for a possibly
more stringent cut-off that includes additonal geometric considerations:\bk
\noindent 
(1) Cosmological paricles absorbed by small black holes: The total number of particles per unit time absorbed by the horizon of a small black hole  is given by
\bea\label{npervolc}
n_c  & = &  \int_{\omega_1 } d\omega  \left[ \left( r_b \omega \right)^2 - \left( {r_c \over r_b } \right)^2 \right]
  N^c_\omega \ , \quad \quad  \omega_1 =\omega_0 = { \kb  \over \epsilon} \\ \nonumber
\eea
The two terms in the square brackets come from the upper and lower limits in the sum over $l$.\bk
\noindent
(2) Black hole particles: The total number of particles emitted by the
black hole per unit time as measured at the cosmological horizon is given by
\be\label{npervolb}
n_b = {A_b \over 4\pi } \int_{\omega_1 } d\omega  \omega ^2  N^b_\omega 
\ee
where 
\be\label{cutoffstwo}
\omega_1 = \omega_0 \simeq {\kb\over  2\pi}  \ \  \ small \ bhs \ , \ \quad\quad \omega_1 \simeq  {1\over l_c^2 \kappa}  \ \  \ large\  bhs
\ee
\noindent
(3) Large black holes: The black hole and cosmological particle production becomes symmetrical with $N^c_\omega \simeq N^b_\omega$ and $n_c \simeq n_b$ with
\be\label{largenpervol}
n_h = {A_b \over 4\pi } \int_{\omega_1 } d\omega  \omega ^2  N^h_\omega \ \ ,
\quad\quad \omega_1 \simeq  {1\over l_c^2 \kappa}  
\ee

\subsection{Particle spectra and total production rates}

For ease of comparison, we repeat the formula for the black hole particle spectrum (\ref{Nbgenone}) derived using 
geometric optics,
\be\label{Nbgenagain}
N^{b}_\omega  =  {2\over \pi \omega R}{\K \over \kc }
\exp [-{1\over {\cal K} }(\omega^{\kb}  \omega_0^{\kc}  )^{1/\kt}   ]  \ , \quad\quad \omega> \omega_0
\ee
The cosmological particle spectra is obtained by interchanging $\kb$ and $\kc$ which gives
\be\label{Ncgenone}
N^{c}_\omega  = {2\over \pi \omega R}{\K \over \kc } \exp [-{1\over {\cal K} }(\omega^{\kc}  \omega_0^{\kb}  )^{1/\kt}   ] 
\ , \quad\quad \omega> \omega_0
\ee
where 
\be\label{Kagain}
  \K =   { \kc\kb \over 2\kt  sin(\pi \kc /\kt ) } 
  \ee
(see (\ref{Kdef}) for the limiting behavior of $\K$). The results of Section (\ref{thermamps}) imply that
it is a good approximation to use
these geometric optics expressions with the appropriate cut-off $\omega_0$ 
for $N_\omega^h$ and $\omega_1$ for $n_h$, given in equations (\ref{npervolc}) to (\ref{largenpervol}).
The total black hole particle production rates is gotten by substituting $N^b_\omega$
into (\ref{npervolb}), 
\be\label{nbgen}
n_b =\frac{A_b \omega_0^{-2\kc /\kb}}{ 4 \pi^2  R sin(\pi\kc/\kt)}
{1\over \K^{2\kt / \kb} }
\Gamma(\frac{2\kt}{\kb},{\omega_0^{\kc / \kt} \omega_1^{\kb / \kt} \over \K } ) 
\ee
where  $\Gamma( s, q) = \int_{q } dy \  y^{s-1} exp[-y]$ is the incomplete Gamma function.
The production rate of cosmological particles is gotten by interchanging the labels $b$ and $c$, giving
\be\label{ncgen}
n_c   =  \frac{A_b \omega_0^{-2\kb /\kc}}{ 4 \pi^2  R sin(\pi\kc/\kt)}
{1\over \K^{2\kt / \kc} }
\Gamma(\frac{2\kt}{\kc},  { \omega_0^{\kc / \kt} \omega_1^{\kb / \kt} \over \K } )
\ee
Equations (\ref{Nbgenagain}) through (\ref{ncgen}) give the spectra and total rate of particle production from the black hole and cosmological horizons, as defined with respect to freely falling frames crossing the cosmological and black hole
horizons respectively, and constitute one of the main results of this paper. An important feature is that in general the spectra
are not thermal.
When applying these formulae one needs to use the appropriate cutt-off frequencies listed in (\ref{npervolc}) through
(\ref{largenpervol}). Explicit formulae for the cut-off are given
 in the limits of small and large black holes.
To find the cut-off $\omega_0$ for a general black hole further analysis of the potential $V$ is needed,
which we defer to future work. Next we focus on the two limits.

\underline{Large black holes:}

In the large black hole limit the mass is going to its maximum value, both horizon areas approach the value
$4\pi l_c^2/3 $,  the surface gravities are going to zero,
$\kappa= \kb\rightarrow\kc\rightarrow 0$, and
 the two spectra appoach the common form
\be\label{Nlarge}
N^{c}_\omega  \simeq N^{ b}_\omega   = 
{1 \over 2 \pi \omega l_c^2 \kappa } exp[  - {4 \omega^{1/2} l_c^{1/4} \over \kappa^{1/4}  } ]
\ , \quad\quad \omega > {1\over l_c^2 \kappa }
\ee
where $R=r_c - r_b \simeq l_c^2 \kappa$ has been used. 
The spectrum is non-thermal, and has the same $\sqrt{ \omega } $ dependence
 in the exponent as was found for $Q=M$ black holes in de Sitter \cite{Kastor:1993mj}, 
 in which case the cosmological and black hole temperatures
 are also equal but non-zero.
 The restiction to higher frequencies comes from (\ref{largenpervol}) and
 the contributing frequencies are in the tail of the distribution. Hence production is highly suppressed.

 Likewise, the total  number of particles produced per unit time goes to zero 
 very rapidly as $\kappa$ goes to zero.
\be\label{ntotbig} 
n_b \simeq n_c \simeq {1\over 6\pi l}{ e^{- 4 /( l_c \kappa )^{3/4}   } \over (l_c \kappa )^{9/4} } 
 \ee
 To find the differences in the two spectra one needs to include corrections to the temperatures.
So large SdS black holes are in a quasi-equilibrium state with the particle absorption by each horizon almost 
balancing its emission, and both processes exponentially suppressed.

\underline{Small black holes:}
Small black holes are defined by the condition that 
  \be\label{epsilon}
  \epsilon = \kc /\kb \ll 1
  \ee
 In this regime, $r_b\simeq 1/(2\kb ) , \ \kc\simeq 1/l_c $, and $R\simeq  l_c $. For particle production by the black hole
  the cut-off frequency is $\omega_0 = \kb /(2\pi )$ and (\ref{Nbgenagain}) becomes
\be\label{Nbsmall}
N^{b}_\omega  = {1 \over \pi^2 \omega l_c } \kb e^{-2\pi\omega/ \kb }
\ee 
which is the expected (high frequency part of a) thermal spectrum for the black hole emission at temperature
 $T_b = \kb / 2\pi$. One might wonder why one does not get the exact Bose-Einstein spectrum, a point that we return
to at the end of this section. Next, 
the total number of particles emitted per unit time (\ref{nbgen}) reduces to
\be\label{smallbtot}
n_b \simeq \frac{A_b }{16 \pi^5 \kc R}   \kb^3    \simeq \  {\kb \over 16 \pi^4 }
\ee
Note that the first expression for $n_b$ has the standard dependence on area and temperature
 for thermal radiation of entropy  by an emitter of area $A$ and temperature $T$,  $n \propto AT^3$, 
 and that the emitted energy
goes like $AT^4$.  The last expression for $n_b$ in (\ref{smallbtot})
 has a non-standard temperature dependence because
the area of a black hole has the non-standard property that it depends on temperature.

Turning now to the cosmological particles,
in the small black hole limit  the cut-off  (\ref{npervolc}) is
 $\omega_0 =\kb /(2\pi \epsilon ) \gg \kb /(2\pi )$, so (\ref{Ncgenone}) becomes  
\be\label{Ncsmall}
N^{c}_\omega  = {1 \over \pi^2 \omega}e^{-y} \ \ , \quad
 y(\omega )\simeq {1\over \epsilon } (\epsilon l_c \omega )^\epsilon \rightarrow {1\over \epsilon } (l_c \omega )^\epsilon
\ee
The spectrum has a slow but non-zero exponential decay,
which is critical to have a finite total particle production $n_c$.  The small black hole limit of $n_c$ in
(\ref{ncgen}) is\footnote{ One finds that a higher order expansion of $N^c_\omega$ is needed if the 
$\epsilon\rightarrow 0$ limit is taken before integrating.}
\be\label{smalltotc}
n_c \sim  {A_b \kb^2 \over  16 \pi^2 l_c \pi}    ( \epsilon)^{ 2/ \epsilon }
\Gamma(\frac{2}{\epsilon},\frac{1 }{\epsilon} )\ =\ { e^{-\kb / \kc }\over 8 \pi^4(\pi-1) l_c } 
\ee
where the asymptotic form of the incomplete Gamma function
 $\Gamma (s, q ) \simeq q^s e^{-q} /(s(\pi -1 ) ) $ for $q\gg 1$ has been used.
Not surprisingly, the absorption of cosmological particles by a very small black hole is very small.

 \underline{Small black holes following Hawking:} 
 
 For small black holes one would expect to recover the Planck spectrum plus corrections, rather than
 just the high frequency tail found in (\ref{Nbsmall}). For small black holes
 $\kb$ gets arbitrarily large, and one might worry that the stationary phase approximation is not
 accurate since it assumes that the frequencies are the largest scales. So as a check
 we return to the integral (\ref{alphab}) for $\alphab$ and using $ \epsilon = {\kc \over \kb} \ll 1$
 expand $ x^{-\epsilon} \simeq (1- \epsilon ln x )$ in the phase factor. This
puts the integrand into a form where the leading term in the $\epsilon$ expansion
 matches the Schwarzchild black hole studied by Hawking \cite{Hawking:1974sw}
\bea\label{alphasmall} 
 \alpha^b_{\omega \omega'} &\simeq & {e^{-i\omega / \kc } \over 4\pi \kb \sqrt{\omega\omega'} }
\int_0^{\infty} dx\left( \omegap +{\omega \over x^{1+\epsilon} } \right) x^{i\omega /\kb } e^{-i\omegap x/\kb } \\ \nonumber
& = & {e^{-i\omega / \kc } \over 4\pi  i \kb \sqrt{\omega\omega'} }
 \left( { i \omegap \over \kb} \right)^{-i\omega / \kb} \left( \Gamma ( 1+{ i\omega \over \kb} ) +
  \left( {i\omegap \over \kb } \right)^\epsilon  \Gamma ( 1- \epsilon + { i\omega\over \kb} ) \right)
\eea
When $\epsilon =0$ this reduces to the result for $\alphab$ for asymptotically flat black holes. 
Analytically continuing in $\omegap$ then gives 
\be\label{betasmalltwo}
 \beta^b_{\omega \omega'}\simeq -i e^{-\pi \omega /\kb } {e^{-i\omega / \kc } \over 4\pi \kb \sqrt{\omega\omega'} }
 \left( {i \omegap  \over \kb } \right)^{-i\omega / \kb} \left( \Gamma ( 1+{ i\omega\over \kb} ) +
  e^{-i \epsilon \pi} \left( {i | \omegap | \over \kb } \right)^\epsilon  \Gamma ( 1- \epsilon + { i\omega \over \kb} ) \right)
\ee

Computing the norms of $\alphab$ and $\betab$ shows 
 that the ratio is unchanged from the Schwarzchild case through leading order in $\epsilon$,
\be\label{ratio}
| \betab |^2 / | \alphab |^2 = e^{-2\pi \omega/ \kb } [1 + {\cal O} (\epsilon ^2 ) ]
\ee
Then the normalization property (\ref{norm}) of the Bogoliubov coefficients  gives
\be\label{spectrumsmall}
N_{\omega}^{b}  \simeq {\gamma_\omega (\epsilon ) \over e^{2\pi \omega / \kb } -1 } 
\ee
where $\Gamma_\omega (\epsilon )$ is the classical absorption cross section of the black hole 
in the SdS spacetime. This elegant result follows because the ratio of the norms 
(\ref{ratio}) is $independent$ of $\omega'$, so (\ref{norm}) reduces to an $\omega$-dependent prefactor times
$N_{\omega}^{b} $. However, this property does not hold for general black holes, so one has to do the integral
over $\omega'$ to get the spectrum.

\subsection{ Horizon fluctuations and the Schottky anomaly}

At the start of this paper we discussed features of the Schottky anomaly for SdS black holes  \cite{Dinsmore:2019elr},
which motivated our choice of
modes that are well-behaved on the horizons to  define particles. What has been learned about quantum
fluctuations in the high and low black hole temperature limits? The large black hole case with $T_b = \kb /2\pi \rightarrow 0$
is the simplest since it is a quasi-equilibrium. Equations (\ref{Nlarge}) and
(\ref{ntotbig}) show that the emission is greatly suppressed, and
going the next step to compute the energy $  E^\phi_b$ and entropy $  S^\phi_b$ in the field particles 
gives \cite{Dinsmore:2019elr}

\be\label{energylarge}
  E^\phi_b \simeq T S^\phi_b \simeq 
  { 1 \over 192\pi^2  l_c} {1 \over ( 2\pi l_c T ) ^{11/4} }  e^{- 4 /(2\pi l_c T )^{3/4}   } 
\ee
and similiarly for the cosmological particles.
Due to the exponential suppression, the derivatives of $n,   E^\phi_b$ and $  S^\phi_b$ with respect to $T_b$
go to zero as $T_b$ goes to zero. Hence we have shown the fluctuations in energy and entropy due to particle production
 share the behavior of classical gravitational fluctuations in the low temperature limit.

The $T_b\rightarrow \infty$ limit is more complicated because of the Schwarzchild black hole instability, which
persists with nonzero $\Lambda$. As is apparent from the
increasing number of particles produced  as the black hole area
goes to zero in (\ref{smallbtot}), the entropy and energy of the produced particles goes to infinity,
\be\label{energysmallb}
E^\phi_b = {3 T_b \over 4}   S^\phi_b \simeq {  l T_b^2 \over 27\pi^3 }
\ee
However, a small black hole carries only a small 
amount of mass and so the emission cuts off after a short amount of time due to back-reaction, which
 must be taken into account. Using arguments analogous to
the asymptotically flat case, but replacing future null infinity with $\fch$, one finds that
in Kruskal time $\Delta U_c$ the temperature increases by an amount
 $\Delta T_b$  \cite{Dinsmore:2019elr} 
 \be\label{ut}
\Delta T_b  = {2\over \pi^3 }T_b^4 \Delta U_c 
 \ee
 and that the energy and entropy in produced particles crossing $\fch$ are  \cite{Dinsmore:2019elr} 
 \be\label{eradu}
\Delta E^\phi_b = {1 \over 32 \pi T_b^2 } \Delta T_b \ ,
\quad\quad  \Delta S^\phi_b =    {1 \over 24\pi T_b^3 } \Delta T_b 
 \ee
 The results (\ref{eradu}) are essentially Schwarzchild physics, except that in the
asymptotically flat spacetime $\Delta U_c$ in (\ref{ut})
 is replaced by the change in a null time coordinate at future null infinity. 
The high temperature limit is complicated because the system is far from equilibrium. The evaporation time scale
is fast compared to $r_b$ and $l_c$, so one needs to look at these dynamical changes in the energy and entropy
 as $T_b$ changes, and see that they go to zero. Hence energy and entropy fluxes due to particle production 
are suppressed at high and low temperatures, as is the case for classical fluctuations. 
 It would be interesting to understand for what size black hole
 the quantum fluxes peak, which is a topic for future work.

\subsection{What about temperatures?}
 One of the striking features of the paticle spectra is that they are not thermal, except for the
 emission from small black holes. Is this result correct? Recall that both the in and out particle states have been
  defined with respect to freely falling observers crossing
 each horizon. On the other hand, if appropriately accelerating observers, namely those who use the Killing time coordinate,
 are used to define the late time particles then a thermal spectrum does result \cite{Bhattacharya:2018ltm}. These
 observers have a special status because their time direction is a symmetry direction. However since they are accelerating,  they must be equipped with ``rockets" to follow their worldlines. This is the case for Rindler observers in Minkowski
 spacetime who measure a thermal spectrum with temperature proportional to their acceleration,
 and that calculation is certainly
 correct. But the Rindler thermal spectrum
  results from an extra force that provides the acceleration, rather than reflecting an intrinsic feature of 
 the spacetime geometry. The spectra calculated in this paper describe the particle production measured for observers
 defined in a natural way near the horizons.
 The analysis of section (\ref{thermamps}) shows that the non-thermal amplitudes
  can be expressed as a convolution of two thermal amplitudes, one at temperature $ \kb /2\pi$ and one at temperature
  $\kc /2\pi$.

Beyond the fact that $T_b$ and $T_c$ naturally appear in the first laws for SdS (\ref{firstm}) and (\ref{firstbetween})
 \cite{Dolan:2013ft}, which extends to 
slow roll inflation \cite{Kim:2014zta} \cite{Gregory:2017sor}  \cite{Gregory:2018ghc},
many analyses of the concept of temperature(s) in SdS have been explored in the literature. Researchers
have studied  several definitions of effective temperature, the role of greybody 
factors, and different choices of observers \cite{Choudhury:2004ph}\cite{Myung:2007my}
 \cite{Urano:2009xn}\cite{ChangYoung:2010ps} \cite{Kim:2014zta}
\cite{Ishwarchandra:2014jca} 
\cite{Hajian:2016kxx}\cite{Li:2016zca}
\cite{Pappas:2017kam}
\cite{Bhattacharya:2018ltm}.
References \cite{ChangYoung:2010ps} and \cite{Pappas:2017kam} attach thermal significance to the radius at 
which $df/dr =0$, the former using holographic arguments.

While there is no contradiction in the fact that the spectra calculated in this paper are not thermal, there is a puzzling
issue concerning the connection between our particle production
results and the temperatures that show up in the first laws (\ref{firstm}) and (\ref{firstbetween}) for SdS. These laws 
relate gravitational perturbations on two different geometrically defined boundaries.
The causal diamond first law (\ref{firstbetween}) 
connects an intersection of the black hole horizon to a space-like related intersection of the cosmological horizon and
relies on the existence of the static Killing field. The surface gravities emerge as part of the structure of the horizon
geometry. On the other hand,
the particle production calculation is done for a scalar field which is added to the system, and no
metric perturbations have been included. The first laws do not imply that perturbative matter fields will exhibit thermal
properties; and yet one might expect that they ``should". 
A concrete way of proceeding to elucidate these differences is to note that
the particle production process is a measurement
on modes that intersect differnt horizons at locations that are time-like or null related. It could be, for example, that
the surface gravity temperatures emerge in correlation functions  computed in our state (defined by
Kruskal-modes)  when the two points are taken close to the black hole horizon
or close to the cosmological horizon. This is another topic for future study.

\section{Conclusions and future directions}\label{conclusion}
We have calculated the spectra and integrated fluxes of produced particles crossing
 each of the future black hole and future cosmological
horizons. Near each horizon particle states are defined with respect the null geodesic, or Kruskal, coordinate there.
As a result the particle fluxes are well-behaved on the horizons, but in general the spectra are not thermal. Interestingly, the amplitude for each spectrum can be written as a convolution
of two $thermal$ amplitudes, one at the cosmological temperture and one at the black hole temperature, 
weighted by the transmission coefficient for wave propagation in the static SdS coordinates. 
The convolution integral also leads to the useful result that the geometric optics approximation is a good one if used
with a low frequency cut-off determined by where the transmission coefficient changes from zero to one.
In the small black limit one recovers a thermal spectrum for the black hole at temperature $\kb /2\pi$, and 
there is a tiny flux of cosmological particles into the black hole. The large black hole limit is a quasi-equilibruim
situation as both temperatures approach the common value of zero and the particle spectra become the same.
One finds that the emission derives only from the tail of the distribution and so
 is exponentially suppressed. As a result,
  the quantum fluctuations in energy and entropy have the property that $\Delta E^\phi / \Delta T_b $
and $\Delta S^\phi / \Delta T_b $ go to zero as $T_b$ goes to zero. This was one of the conditions needed for 
the behavior of quantum fluctuations on the horizon to be consistent with the Schottky anomaly behavior of classical
gravitational fluctuations.
 In the small black hole limit, we have argued that after
taking into account the dynamical nature of the rapidly evaporating black hole and the small total energy
available, one also finds  that $\Delta E^\phi / \Delta T_b $ and $\Delta S^\phi / \Delta T_b $ go to zero as
  $T_b \rightarrow \infty$.

There remain many questions. It would be interesting to
understand the emission and absorption for a generic black hole. How big does a black hole have to be to
see the effects of $\Lambda$? At what size is the net black hole emission slowed down by a significant factor?
Further analysis of the transmission coefficient is required to answer these questions. In this paper we have only used
the general feature of $|T_{\nu l} |$. But there could be
interesting physics in its behavior at the frequencies picked out by the point of stationary phase. It is also important
 understand further of the properties of the state used here, including its two-point function and 
stress tensor, building on the work of \cite{Bates:2014jba}\cite{Aalsma:2019rpt} to include a black hole, and
 extending the calculations of \cite{Markovic:1991ua}\cite{Choudhury:2004ph} to beyond the cosmological horizon.

In the times of precision cosmology, one should ask if the signature of primoridal black holes during an epoch
of inflation could be observed in the cosmic microwave background. A detailed calculation addressing this question was done in \cite{Prokopec:2010nm} using a different state. It would be interesting to 
 extend our calculations to include the behavior of  modes in the far field
beyond the cosmological horizon, as well as the time dependence of the mass and effective $\Lambda$ in 
slow-roll inflation \cite{Gregory:2018ghc},
 and apply these results to the calculation of quantum fluctuations in inflation.
.

\subsection*{Acknowledgements} 
We would like to thank Paul Anderson, Patrick Draper, David Kastor, and Lorenzo Sorbo for multiple useful
 conversations.
We would also like to thank NORDITA for their hospitality and support during
the {\it Cosmology and Gravitational Physics with Lambda} Scientific Program August 2018, and the
Centro de Ciencias de Benasque Pedro Pascual for their hospitality during the 
{\it Gravity-New Perspectives from Strings and Higher Dimensions} workshop July 2019.
Y.Q. was partially supported by the US-NSF grant PHY-1820675.

\vskip 0.2in

\appendix

{\bf APPENDIX}\label{appendix}
\section{Deriving normalized spectra and rates}\label{discrete}

To get properly normalized expressions for the quantum spectra and the integrated rates of particle production
one needs to use properly normalized wave packets. 
In terms of the following
steps this is equivalent to putting the system in a box of size $R^3$ so that the allowed frequencies are integer indexed
$\omega_j = j/R$.
Since it is not clear what it means to put an SdS black hole in a box, we prefer to think about
using normalizable states. 
The key results are equations (\ref{numberopsthree}) and (\ref{rate})
for the particle spectra and the total number of particles produced per unit time expressed in terms of the continuous basis
functions, and which depend on the scale $R$.

To derive these results,
let $F_j $ and $P_j$ be normalized
wavepackets built out of the $f_\omega$ and the $p_\omega$ basis modes respectively, so that the packet is
 peaked at frequencies near $\omega_j = j /R $ [ref-Hawking], where $R$ is a length scale, and normalized as
\be\label{innerprod2}
(F^h_j , F^h_k ) = \delta _{ij} \  ,\quad and \quad (P^h_j , P^h_k ) = \delta _{ij} 
\ee
The associated creation and annhilation operators satisfy $[a^{h\dagger}_j , a^h_k ]
= [b^{h\dagger}_j , b^h_k ] =\delta_{jk}$, and the field is expanded as
\be\label{phi2}
\Phi =\sum_{jlm} [ P^b_{ \ j l} a^b_{\ jl} + P^c_{\ jl} a^c_{\ jl} + h.c. ] \  = \sum_{jlm} [ F^b_{\  jl}  b^b_{\ jl} + 
F^c_{\  jl} b^c_{\ jl} + h.c. ]
\ee
Then the two spectra are given by
\begin{eqnarray}\label{numberops} 
 N_j^b & = &
<0| \  b_j^{c\dagger}b^c_j \  |0> \ \ ,\quad emission\ from \ black \ hole \nonumber\\
N_j^c & = &
<0| \  b_j^{b\dagger} b_j^b \  |0> \ \ ,\quad emission\ from \ cosmo \ horizon \nonumber\\
\end{eqnarray}
The quantities $N_j^h$ are dimensionless.
Note that the odd-looking mismatch of labels between the left and right sides of these equations is because, for example,
the particle emission interpreted as coming $from$ the black hole is measured with the particle operators that
are far away from the black hole, in this case the cosmological horizon. The notation ``rights itself" in our subsequent
expressions (\ref{numberopstwo}) for the $ N_j^h$ in terms of the Bogoliubov coefficients.

 We need to convert the
relation (\ref{numberops}) for particle spectra to the basis of continuous functions.
Since the commutation relations of the creation and annhilation operators have different dimensions in the two cases,
the dimensions of the operators differ, and hence the dimensions of the mode functions  and the 
Bogoliubov coefficients change as well. The changes in dimensions are summarized by\footnote{These relations 
are impressionistic in the sense that they hold in converting sums to integrals, but equation (\ref{relatedims}) is
easier to read.}
\be\label{relatedims}
 b_j = {1\over \sqrt{R} } b_{\omega_j}  \ , \quad F_j ={1\over \sqrt{R} } f_{\omega_j} \ , 
 \quad  \beta_{jk} = {1\over R} \beta_{\omega_j \omega_k}
\ee
The basis expansions (\ref{expandmodes}), (\ref{expandops}), and the formulas for the Bogoliubov coefficients (\ref{coeffs})
become sums over the basis elements, 
\be\label{expandops2} 
b^c_j= \sum_k \left[ \alpha_{jk}^{b*} a^b_k-
\beta_{jk}^{b*} a_k^{b\dagger} 
+ A_{jk}^{c*} a^c_{k}-
B_{jk}^{b*} a _k^{c\dagger} \right]
\ee
and similiarly for the expansions of the other particle operators. We also note the normalization propertiy of 
the expansion coefficients,
\be\label{norm}
\sum_k \left( |\alpha^b_{jk} |^2 - | \beta^b_{jk }|^2 \right) =
1 - \sum_k \left( |A^b_{jk} |^2 - | B^b_{jk}  |^2 \right)
 =\ \gamma^b (\omega_j )
\ee 
where $\gamma^b (\omega_j ) $ is the classical absorption coefficient of the black hole, often referred to as a greybody
factor. An analogous
relation holds interchanging the indices $b$ and $c$.

Hence equation (\ref{expandops2}) for
the probablity of  horizon $h$ producing a late-time particle with  frequency $\omega_j$, in the early-time
 vacuum state (\ref{vacdef}),  becomes 
 \be\label{numberopsapp}
N_{\omega_j } ^{ h } = \Sigma_k \  |\beta^h_{j k } |^2 \   
  \ , \ \quad h=b,c
\ee
 Using (\ref{relatedims}) and the relation $\sum_k = R\int d\omega' $,
then (\ref{numberopsapp}) becomes
\be\label{numberopstwoapp}
N_{\omega }^{ h } = \ {1\over R} \int d\omega' |\beta^h_{\omega \omega'  } |^2 
\ee
in the continuous basis.
Note the important dimensionful factor of $1/R$ that multiplies the integral using the continuous basis functions, and
that $N_\omega ^{ h } $ is dimensionless. This is the desired result.

\section{Stationary Phase to evaluate $\alphab$}\label{staphaapp}

The Boguliubov coefficient (\ref{alphabexact}) can be written as
 \bea\label{alphad}
 \alpha_{\omega \omega'} 
  &=&\frac{1}{8\pi^2\kb\kc\sqrt{\omega\omega'}} \int_0^{\infty} dx dy \  e^{-i\omega \frac{y}{\kc}} e^{-i\omega' \frac{x}{\kb}}
\int_{-\infty}^{\infty} d{\nu} \  T^*_{\nu} \ e^{i\nu(\frac{ln x}{\kb}+\frac{ln y}{\kc})}H (x,y,\nu)
\eea
where
\bea
H(x,y,\nu)
  &=&\frac{\nu}{xy}
+\frac{\omega}{x}
+\frac{\omega'}{y}
+\frac{\omega\omega'}{\nu} \nonumber
 \eea
 In this appendix we show that two applications of stationary phase bring this multiple integral to the geometric optics
 form (\ref{alphabgo})
with the important difference that the transmission coefficient is included in the integrand.
 A third application of stationary phase plus analytic continuation gives the expression for $\betab$ used to get 
 the particle spectra in section (\ref{stapha}).
 
 The geometric optics approximation motivates the change of variables 
 $z=yx^\lambda$ with $\lambda =\kc /\kb $ in (\ref{alphad}). This gives
 our starting point for using stationary phase,
\be\label{alphalong}
\alphab =  {1\over 8\pi^2 \sqrt{\omega\omega'}\kb\kc } \int_0^\infty dx dz \
e^{ -i  \omega z x^{-\lambda}/\kc}   e^{-i\omega' x/\kb }
 \int_{-\infty}^\infty d\nu T^*_\nu \ H (x,z,\nu )
e^{ i  \nu ln z  / \kc }
\ee
where $H$ is now
\be\label{hdef}
H (x, z, \nu ) =  { \omega' \over z}+{\omega \over x^{1+\lambda} } 
+ {\nu \over zx}  +{\omega\omega' \over \nu x^\lambda }
\ee

We first do the integral over $z$ in (\ref{alphalong}), taking  $\phi (z)=- \omega z x^{-\lambda}   + \nu ln z $.
The point of stationary phase is at $z_0 =  x^\lambda \nu /\omega$ and using (\ref{stapha}) gives
\be
\alphab \simeq  {\sqrt{2\pi\kc} e^{-\frac{i\pi}{4}} \over 8\pi^2 \sqrt{\omega\omega'}\kb\kc } \int_0^\infty dx \
  e^{-i\omega' x/\kb } \frac{x^{\lambda}}{\omega}
 \int_{-\infty}^\infty d\nu \sqrt{\nu} \ T^*_\nu  \ H (x,z_0,\nu ) 
 e^{ i\nu [-1+ln(\frac{x^{\lambda}\nu}{\omega}]) ]/\kc}
\ee
Next, take the phase  to be $\phi (\nu ) =\nu (-1 +ln(\frac{x^{\lambda}\nu}{\omega}) )$. The stationary point is at
 $\nu_0=x^{-\lambda}\omega$, and doing the integral over $\nu$ gives
 \be\label{alphabgen}
\alphab \simeq  {1\over 2\pi \sqrt{\omega\omega'}\kb } \int_0^\infty dx \
    T^*_{\nu_0}  \ (\omega'+\frac{\omega}{x^{1+\lambda}})
 e^{-i\omega' x/\kb } e^{ -ix^{-\lambda}\omega/\kc}
\ee
This integrand is the same as the integrand found using the geometric optics approximation, equation (\ref{alphabgo}), 
with the added ingredient that the transmission coefficient $T_\nu$ is in the integrand. 
 
The third application of stationary phase is the same as that used to derive {\ref{alphab}) giving
\be\label{alphabthree}
\alphab 
 ={\sqrt{2}  e^{-i\frac{\pi}{4}}   \ T^*_{\nu_0} \over \sqrt{\pi} \sqrt{\kt}  ( \omega^{\kc} \omega^{\prime \kb} )^{1/2\kt} }  \
exp[ -i{  \kt \over \kc \kb}   (\omega^{\prime\kc} \omega^{\kb })^{1/\kt }  ]
\ee
where $ \nu_0  = (\omega^{\prime \kc} \omega^{\kb} )^{1/\kt } $. 

Analytically continuing $\omega'  \rightarrow
|\omega' | e^{-i\pi} $ as discussed preceeding equation (\ref{betabgo}) gives $\betab$
\begin{equation}\label{betab}
\beta^b_{\omega \omega'} \simeq 
- i{   \sqrt{2} e^{-\frac{i\pi}{4}} 
 e^{i\pi \kb /2 \kt} \over \sqrt{ \pi  \kappa_T } (\omega^{\kc} \omega^{\prime\kb })^{1/2\kt } } 
T^*_{ \sigma  } \exp [- {\kappa_T \over \kappa_b \kappa_c }
 (\omega^{\prime\kappa_c}  \omega^{\kappa_b}  )^{1/\kappa_T} (i \cos \pi {\kappa_c \over \kappa_T} 
 +\sin \pi {\kappa_c \over \kappa_T} ) ]
 \ee
 where $ \sigma$ is the complex frequency
  \be\label{wstar}
   \sigma = (\omega^{\prime \kc} \omega^{\kb} )^{1/\kt }e^{-i\pi \kc/\kt }
   \ee

\section{Delta function method}
As a check on the stationary phase approximation, we use another way to compute the $ \alpha^b_{\omega \omega'} $ coefficients. The idea is to replace  $T_\nu$ with unity in the exact expression, and then show that 
the difference is subleading. Unfortunately there is one term in (\ref{alphalong}) that can not be evalutated 
by the second method, so this does not fully provide an alternative evaluation.
Rewrite equation \ref{alphad} as
 \bea\label{appalphab}
 \alpha^b_{\omega \omega'} 
  &=&
\frac{1}{8\pi^2\kb\kc\sqrt{\omega\omega'}} \int_0^{\infty} dx dy \  e^{-i\omega \frac{y}{\kc}} e^{-i\omega' \frac{x}{\kb}}
\int_{-\infty}^{\infty} d{\nu} \  e^{i\nu(\frac{ln x}{\kb}+\frac{ln y}{\kc})}
(\frac{\nu}{xy}
+\frac{\omega}{x}
+\frac{\omega'}{y})
\nonumber\\
&&+\frac{\sqrt{\omega\omega'}}{8\pi^2\kb\kc} \int_0^{\infty} dx dy \  e^{-i\omega \frac{y}{\kc}} e^{-i\omega' \frac{x}{\kb}}
\int_{-\infty}^{\infty} d{\nu} \   e^{i\nu(\frac{ln x}{\kb}+\frac{ln y}{\kc})}
\frac{ T^*_{\nu}}{\nu} \nonumber\\
&&+\frac{1}{8\pi^2\kb\kc\sqrt{\omega\omega'}} \int_0^{\infty} dx dy \  e^{-i\omega \frac{y}{\kc}} e^{-i\omega' \frac{x}{\kb}}
\int_{-\infty}^{\infty} d{\nu} \  (T^*_{\nu} -1)\ e^{i\nu(\frac{ln x}{\kb}+\frac{ln y}{\kc})}
(\frac{\nu}{xy}
+\frac{\omega}{x}
+\frac{\omega'}{y})\nonumber\\
&=&\alpha^{(1)}_{\omega \omega'} +\alpha^{(2)}_{\omega \omega'} +\delta\alpha_{\omega \omega'} 
\eea

The first term can be simplified by noting that the integral over $\nu$ gives delta functions,
\begin{eqnarray*}
\alpha^{(1)}_{\omega \omega'} 
&=&\frac{1}{4\pi \kb\kc\sqrt{\omega\omega'}} \int_0^{\infty} dx dy \  e^{-i\omega \frac{y}{\kc}} e^{-i\omega' \frac{x}{\kb}}\left[
  \delta'({\frac{ln x}{\kb}+\frac{ln y}{\kc}})
\frac{1}{xy}
+\delta({\frac{ln x}{\kb}+\frac{ln y}{\kc}})(\frac{\omega}{x}
+\frac{\omega'}{y})\right]\\
&=&\frac{1}{4\pi \kb\sqrt{\omega\omega'}} \int_0^{\infty} dx  \  
e^{ -\frac{i\omega x^{-\lambda}}{\kc}} e^{-i\omega' \frac{x}{\kb}}
\left(
\frac{2\omega}{x^{1+\lambda}}+\omega'
\right)\\
\end{eqnarray*}

This is almost the result we obtained in the geometric optics limit, but needs a factor of $2$ multiplying $\omega'$
in the integrand. Turning to 
 $\alpha^{(2)}_{\omega \omega'} $, one again uses successive applications of stationary phase, which gives
\begin{eqnarray*}
\alpha^{(2)}_{\omega \omega'} 
&=&\frac{1}{4\pi \kb\sqrt{\omega\omega'}} \int_0^{\infty} dx  \ T^*_{\nu_0} \  \omega' 
e^{ -\frac{i\omega x^{-\lambda}}{\kc}} e^{-i\omega' \frac{x}{\kb}}
\end{eqnarray*}

Where $\nu_0=(\omega'^{\kc}\omega^{\kb})^{1/\kt}$. From the analysis in the previous section, we know that since $\omega$ and $\omega'$ are much greater than $\omega_0$, $\nu_0$ is also much greater than $\omega_0$. Hence $T_{\nu_0}^*$ is approximately equal to 1. Then the sum  $\alpha^{(1)}_{\omega \omega'} + \alpha^{(2)}_{\omega \omega'} $ is 
equal to the geometric optics approximation.

 To complete the argument we need to check that the magnitude of $\delta \alpha_{\omega \omega'}$ is 
 small. 
Since $|T^*_{\nu}-1|<1$, $|\delta \alpha_{\omega \omega'}|$, after doing the integral over $\nu$ one has
\begin{eqnarray}
|\delta \alpha_{\omega \omega'} |
&< &|\frac{1}{8\pi^2\kb\kc\sqrt{\omega\omega'}} \int_0^{\infty} dx dy \  e^{-i\omega \frac{y}{\kc}} e^{-i\omega' \frac{x}{\kb}}
\{2i\frac{\sin [\omega_0 (\frac{ln x}{\kb}+\frac{ln y}{\kc})]-\omega_0 (\frac{ln x}{\kb}+\frac{ln y}{\kc}) \cos [\omega_0 (\frac{ln x}{\kb}+\frac{ln y}{\kc})]}{xy(\frac{ln x}{\kb}+\frac{ln y}{\kc})^2 }\nonumber\\
&&+\frac{2sin\left[\omega_0(\frac{ln x}{\kb}+\frac{ln y}{\kc})\right]}{(\frac{ln x}{\kb}+\frac{ln y}{\kc})}(\frac{\omega}{x}+\frac{\omega'}{y})\}|\nonumber\\
&\approx&|\frac{e^{-i \omega_0(\frac{1}{\kc}+ \frac{1}{\kc})} }{ 2 \pi z\sqrt{\kb\kc\omega\omega'}}  \  
\left[\frac{\sin (\omega_0 z)}{\omega_0z }- \cos (\omega_0 z)
-2isin(\omega_0z)\right]|
\label{dealp}
\end{eqnarray}
where stationary phase approximation has been used to do the integrals over $x$ and $y$
 and $z=\frac{1}{\kb}ln (\frac{\omega_0}{\omega'})+\frac{1}{\kc}ln (\frac{\omega_0}{\omega})$.  
So the ratio is approximately equal to
\begin{eqnarray}
\frac{|\delta\alpha_{\omega\omega'}|^2}{|\alpha^b_{\omega\omega'}|^2}
&\simeq &\frac{\kt}{8\pi \kb\kc\omega^{\kb/\kt}\omega'^{\kc/\kt} \ z^2}
\left[\left(\frac{\sin (\omega_0 z)}{\omega_0z }- \cos (\omega_0 z)\right)^2
+4sin^2(\omega_0z)\right]
\end{eqnarray}
For $\omega$ and $\omega'$ much greater than the cutoff, $z\rightarrow -\infty$,  which gives
$\frac{|\delta\alpha_{\omega\omega'}|^2}{|\alpha^b_{\omega\omega'}|^2}
 \rightarrow 0$. Hence the contribution from $\delta \alpha_{\omega\omega'}$ is small compared to $\alpha^{(1)}_{\omega\omega'}$ and $\alpha_{\omega\omega'}^{(2)}$. 
To summarize, replacing $T_\nu$ by one in (\ref{appalphab}), one can evaluate three of the four terms
by Fourier transforms. The fourth term contains the factor $T_\nu /\nu$ which is well behaved as $\nu$ goes to
zero since $T_\nu$ vanishes, so it is not useful to put $T_\nu =1 $ since that introduces an artificial singularity, and
 stationary phase was used instead. The correction term $ \delta\alpha_{\omega\omega'} $ is small compared to 
 the geometric optics result. Note that a more precise analysis of the the low frequency behavior of the transmission
 coefficient could allow a better approximation to the fourth term.

\section{Energy and entropy of produced particles }
The particle spectra and total emisson rates are given in equations (\ref{Nbgenagain}) to (\ref{ncgen}).
The energy fluxes in the particles are given by
\begin{eqnarray}
E_c&=&\int_{\omega_1} d\omega \ \omega^3r_b^2 N_{\omega}^c \nonumber\\
&=&\frac{ 4 r_b^2 }{4\pi R \ sin(\frac{\pi}{\e+1} )}
[\frac{\e\kb}{2(1+\e)sin(\frac{\pi}{\e+1})}]^{\frac{3}{\e}+3} \omega_0^{-\frac{3}{\e}} 
\Gamma(\frac{3}{\e}+3,\frac{\omega_0^{\kb/\kt}\omega_1^{\kc/\kt}}\K )
\end{eqnarray}
and
\begin{eqnarray}
E_b&=&\frac{A}{4\pi}\int_{\omega_1} d\omega  \omega^2  N^b_{\omega}\nonumber\\
&=&\frac{A\omega_0^{-3\frac{\kc}{\kb}}}{ 4 \pi^2 R sin(\pi\kc/\kt)} \K^{\frac{3\kt}{\kb}}
\Gamma(\frac{3\kt}{\kb},\frac{\omega_0^{\kc/\kt}\omega_1^{\kb/\kt}} \K ) \nonumber\\
\label{ebprod1}
\end{eqnarray}
where $\epsilon = \kb / \kc$ is not necessarily small in these expressions.

For small black holes take $\e\ll 1$ limit and use $\omega_0 = \omega_1 = \kb /2\pi$,
\begin{eqnarray}
E_b&\approx&
\frac{\kb^2}{108\pi^5}\Gamma(3,1)
\end{eqnarray}
which gives equation (\ref{energysmallb}).
For large black holes, take the limit $\epsilon \rightarrow 1$ with
 $\omega_0=\kappa^{3/2}l^{1/2}$, $\omega_1=\dfrac{1}{\kappa l^2}$, and $R=\kappa l_c^2$. One obtains
\begin{eqnarray}
E_c\rightarrow E_b =E 
\approx \frac{A \kappa^{1/2} }{4^7 \pi l_c^{7/2}}\Gamma(6,\frac{4}{(\kappa l_c)^{3/4}})\nonumber\\
 \end{eqnarray}
which then gives equation (\ref{energylarge}).
The entropy in the produced particles comes from using the first law $dE =TdS$ \cite{Dinsmore:2019elr}.

\end{document}